%% file: main.tex
\documentclass[sigconf,10pt]{acmart}
\input{macros}

\usepackage{pifont}
\usepackage{enumitem}
\usepackage{xspace}
\AtBeginDocument{%
  }
\usepackage{threeparttable}
\usepackage{xspace}
\usepackage{subcaption}
\usepackage{graphicx}

\usepackage{booktabs}
\usepackage{tabularx}
\usepackage{makecell}
\usepackage{array}
\usepackage{multirow}
\usepackage{subcaption} 
\usepackage{graphicx}

\newcolumntype{L}{>{\raggedright\arraybackslash}X}
\usepackage{xcolor}
\copyrightyear{2026}
\acmYear{2026}
\setcopyright{cc}
\setcctype{by}
\acmConference[MobiCom '26]{The 32nd Annual International Conference on Mobile Computing and Networking}{October 26--30, 2026}{Austin, TX, USA}
\acmBooktitle{The 32nd Annual International Conference on Mobile Computing and Networking (MobiCom '26), October 26--30, 2026, Austin, TX, USA}
\acmDOI{10.1145/3795866.3844465}
\acmISBN{979-8-4007-2505-0/2026/10}
\usepackage{adjustbox}

\usepackage{siunitx}

\newcommand{\cmark}{\textcolor{green}{\checkmark}}
\newcommand{\xmark}{\textcolor{red}{\ding{55}}}
\newcommand{\sysname}{\textsc{RadioSight}\xspace}

\usepackage[font=small,labelfont=bf]{caption}

\begin{document}
\title{RadioSight: Predictive mmWave XR Network Optimization from Dynamic Neural Radio Fields}

\settopmatter{authorsperrow=4}
\author{Lihao Zhang}
\affiliation{\institution{University of Georgia}\country{}}
\author{Paul Kudyba}
\affiliation{\institution{University of Georgia}\country{}}
\author{Zhenlin An}
\affiliation{\institution{University of Georgia}\country{}}
\author{Haijian Sun}
\affiliation{\institution{University of Georgia}\country{}}

\begin{abstract}
Next-generation extended reality (XR) networks rely on mmWave communication for multi-gigabit throughput, yet highly directional links are vulnerable to user mobility and blockages, causing frequent outages under reactive beam management. Emerging neural radio fields can predict radio propagation, but prior work remains limited to offline channel reconstruction. We introduce \sysname, a real-time multi-modal radio field system for predictive mmWave optimization and proactive Multi-User MIMO beamforming. \sysname combines backward beam-tracing with real-time semantic object synchronization to anticipate RF geometry changes without full model retraining. Implemented as an edge-executable pipeline for commercial 28\,GHz arrays, \sysname determines each scheduling window's beams during the preceding window without exhaustive beam sweeps. Experiments show that \sysname reduces beam-search error by up to $\sim$50\%, improves median throughput by $2\times$, and enhances link stability. 
\end{abstract}



\maketitle

\section{Introduction}
\label{s:intro}
Millimeter-wave (mmWave) communication is a cornerstone of next-generation augmented reality (AR), virtual reality (VR), and extended reality (XR) networking, delivering multi-gigabit throughput with highly directional beamforming~\cite{sur2017wifi,abari2016cutting}. This capacity underpins bandwidth-intensive immersive applications such as volumetric video streaming~\cite{zhang2022yuzu}, cloud-based XR gaming~\cite{mangiante2017vr}, multi-user virtual collaboration~\cite{liu2024muv2}, and AR-assisted remote surgery~\cite{simsek2016tactile}. As XR devices evolve toward always-on, mobility-rich usage, reliable mmWave connectivity becomes increasingly critical.

\begin{figure}[t!]
    \centering
    \includegraphics[width=0.95\linewidth]{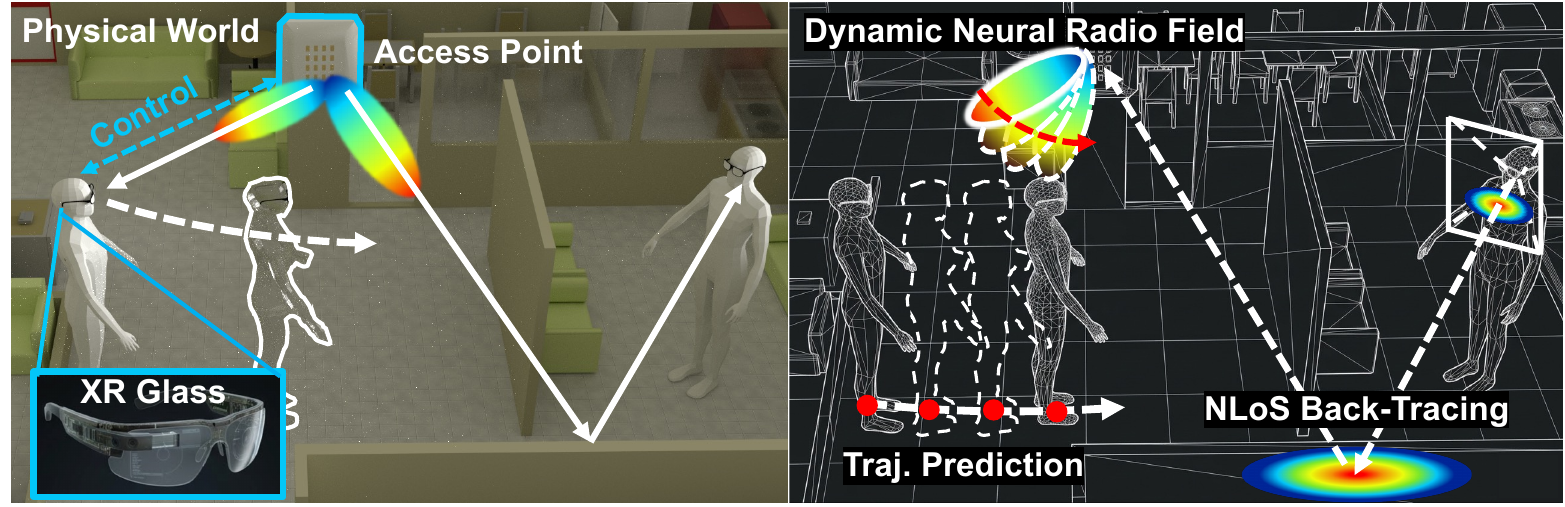}\vspace{-4mm}
    \caption{System Overview of \sysname. \textnormal{(Left) \textbf{Physical World:} \sysname anticipates impending link failures by forecasting the XR user's future trajectory (dashed arrow). (Right) \textbf{Neural Radio Field:} Rather than reactively scanning after a disconnection, the system queries its continuously updated radio digital twin. By synthesizing the spatial spectrum at predictive viewpoints, \sysname proactively discovers optimal non-line-of-sight (NLoS) reflection paths to execute seamless, zero-downtime beam switching.}}
    \vspace{-4mm}
    \label{fig:teaser}
\end{figure}

Delivering on this promise, however, is hindered by a fundamental fragility: mmWave links are highly susceptible to user motion~\cite{zhang2024habitus}, head rotation~\cite{woodford2021spacebeam}, and environmental dynamics~\cite{jain2024commrad,chen2024rfcanvas}. 
Because even minor spatial shifts can invalidate an optimal beam, traditional reactive approaches~\cite{jain2021two,hassanieh2018fast} trigger exhaustive beam scanning upon disconnection. This introduces long-tail latency that severely degrades immersive XR experiences. To preempt such disruptions, neural network (NN) systems~\cite{zhang2024habitus,narayanan2020lumos5g,jain2021two,xu2025roaming} predict signal drops using historical link quality and device telemetry. However, by treating the environment as a temporal ``black box,'' these methods lack a foundational understanding of the physical space. Consequently, while they can predict \textit{that} a link will fail, they cannot explain \textit{why}, identify alternative propagation paths, or explicitly adapt to ongoing dynamics.

Recently, the emergence of Neural Radio Fields~\cite{zhao2023nerf2,zhang2026rf,lu2024newrf,yang2024rnerf,umer2025neural,hoydis2023sionna,an2025radiotwin} has opened a promising avenue by providing continuous, high-fidelity models of wireless propagation at arbitrary 3D positions. Unlike traditional black-box predictors, a radio radiance field captures the \textit{physical reality} of the environment (e.g., structural geometry and multipath propagation characteristics). This understanding makes it possible not only to predict link quality but also to \textit{explain} signal degradation and \textit{identify} optimal alternative paths. Ultimately, this capability unlocks a fundamentally new paradigm for mmWave optimization: rather than reacting to observed disconnections, a network controller can leverage a deterministic understanding of the radio environment to \textit{proactively} steer beams before blockages even occur.

However, most existing neural radio field studies~\cite{zhang2026rf,zhao2023nerf2} focus on static, site-specific, offline channel reconstruction rather than maintaining a real-time dynamic radio twin. Although recent systems such as RFCanvas~\cite{chen2024rfcanvas} attempt to track environmental dynamics by fusing vision and RF observations, their reliance on computationally heavy implicit models and complex retraining limits their suitability for latency-sensitive networking applications.  Moreover, these approaches only recover the receiver-side spatial spectrum and cannot infer the transmitter-side spatial spectrum required for downlink beamforming. Therefore, they lack the architectural integration required to translate volumetric radio intelligence into instant, actionable network decisions, such as beam steering or link handover. Consequently, a critical gap remains between high-fidelity channel modeling and practical, system-level predictive control.

To address the limitations of existing systems (Table~\ref{tab:system_comparison}), we propose \sysname, the first practical system that transforms a radio radiance field into a real-time predictive mmWave optimization engine (Fig.~\ref{fig:teaser}). Our key insight is that XR platforms already provide a rich sensing infrastructure (camera and IMU) that can be leveraged to continuously capture environmental dynamics. By fusing these visual observations with radio measurements, \sysname constructs a multi-modal radio twin using an extremely efficient 3D Gaussian Splatting (3DGS) representation. This design enables the system to maintain a continuously updated, physics-guided radio field that reflects both static geometry and dynamic objects, allowing the radio environment to be synchronized with real-world changes, through lightweight edits rather than full retraining that prior radio fields require.

\begin{figure}[t!]
    \centering
    \includegraphics[width=0.45\textwidth]{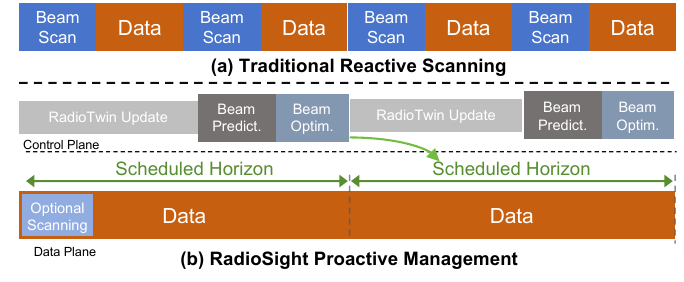}\vspace{-5mm}
    \caption{Comparison of Beam Management Strategies. \textnormal{Top: Traditional Reactive Scanning interrupts data for frequent beam sweeps. Bottom: \sysname decouples beam management onto the control channel, maintaining continuous transmission. The optional scan is a sparse, asynchronously triggered operation.}}\vspace{-6mm}
    \label{fig:system_timeline}
\end{figure}

Building on this real-time radio twin, \sysname enables proactive network optimization. 
Specifically, the system predicts user trajectories and queries the dynamic radio field to estimate future channel states, allowing it to identify optimal beam configurations ahead of time. As illustrated in Fig.~\ref{fig:system_timeline}, \sysname operates over continuous scheduling horizons: to mask computational latency, beam management for the upcoming horizon is executed during the preceding one. During this period, the system rapidly synchronizes the radio field with environmental changes (e.g., user motion or transient blockages), and then performs predictive beam optimization based on the updated model, enabling seamless communication even under rapidly changing conditions. The demo videos, open-sourced code and datasets of dynamic multi-modal neural radio field are on \href{https://pervasive-intelligence-lab.github.io/radiosight/}{\textcolor{blue}{https://pervasive-intelligence-lab.github.io/radiosight/}}. 

However, realizing this proactive vision requires addressing three key technical challenges:

\textit{First, constructing a multi-modal radio twin to support backward beam-tracing.} Traditional radio radiance fields implicitly encode RF properties, leading to geometric ambiguity and difficult local editing that hinders downstream tasks. \sysname overcomes this by constructing a multi-modal radio field that fuses visual features to enforce geometric fidelity and semantic features to segment movable objects. Leveraging this accurate surface representation, we introduce a back-tracing mechanism that traces multipath arrivals from the receiver back to the transmitter. This explicitly identifies the optimal angle of departure (AoD) for proactive beamforming without exhaustive channel sounding.

\textit{Second, achieving real-time twin synchronization under environmental dynamics.} A major bottleneck in maintaining a ``live'' radio twin is the computational overhead of full-model retraining when the environment changes. To address this, \sysname introduces a highly efficient, two-tiered update mechanism built upon the explicit radio 3DGS architecture. Rather than aggressively updating the field for transient blockages, \sysname leverages a novel co-rendering approach: it tracks transient objects using lightweight 3D bounding boxes and co-renders their impacts with the static radio field to instantly mask out blocked radio radiance. For long-term environmental changes involving movable objects, the system performs incremental updates by tracking only the displaced objects and specifically adjusting their corresponding radio appearance. This surgical synchronization strategy maximizes real-time responsiveness with minimal computational cost.

\textit{Third, proactive cross-layer network orchestration to combat user mobility.} When the user moves, reactive beam management is fundamentally limited by feedback loop latency: even with instantaneous pose measurement, the round-trip delay of executing a beam switch exceeds the mmWave channel coherence time. To eliminate this bottleneck, \sysname predicts how the user will move in the future based on client-side SLAM and kinematic models. By pre-querying the synchronized radio twin using these future trajectories, the system anticipates impending blockages and discovers optimal alternative paths ahead of time. This enables a proactive, cross-layer control system that seamlessly schedules beam switching and rate adaptation before the user experiences any actual link degradation.

\textbf{Contributions.} We implement \sysname as a full-stack, edge-executable pipeline integrated with a commercial 28\,GHz testbed and Meta Aria glasses~\cite{meta2024aria}. By systematically co-designing 3DGS radio fields~\cite{zhang2026rf}, vision-language understanding~\cite{qiu2024feature}, and mobility prediction~\cite{weng2020ab3dmot}, \sysname enables a practical dynamic neural radio field for real-time XR network optimization. Through extensive real-world evaluations, we summarize our core contributions:

(1)~We propose a new real-time, editable dynamic neural radio field by using semantic information to anchor 3DGS segmentation and drive concise updates, supported by a distributed UE-feedback architecture.

(2)~We introduce backward beam-tracing. Unlike prior radiance-field approaches, we use accurate scene geometry to trace radio paths backward, inferring dynamics-aware optimal beams for the transmitter.

(3)~We build proactive network optimization atop this predictive field, pre-emptively switching paths before blockages or LoS-to-NLoS transitions occur to prevent disconnections.

(4)~Extensive real-world evaluations in diverse XR mobility scenarios show that \sysname substantially reduces beam alignment errors by up to $\sim$50\%, improves median throughput by up to 2$\times$, and eliminates the rescans and disconnections inherent to reactive schemes.

\vspace{-2mm}\section{Related Work} \label{sec:related_works}
\sysname spans many active research topics in radio fields and mmWave XR Networks.

\textbf{Neural Radio Fields.} Neural representations for radio environment modeling have advanced rapidly. Ray tracing~\cite{woodford2021spacebeam}, NeRF-based methods~\cite{zhao2023nerf2, lu2024newrf, an2025radiotwin} synthesize radio signals across 3D space, while 3DGS-based approaches~\cite{zhang2026rf, wen2026neural, yang2026gsrf} offer faster reconstruction. However, these methods are designed for \textit{receiver-side} spatial spectrum prediction and cannot derive the transmit-side AoD needed for MU-MIMO beamforming, particularly under NLoS reflection paths. 
Moreover, these approaches tackle \textit{static} environments and cannot adapt to dynamic occlusions without costly retraining.
A parallel line of work leverages ray-tracing simulations~\cite{hoydis2023sionna, cao2024raypronet, ma2024automs} for site-specific coverage prediction and deployment optimization, but these remain offline planning tools unsuitable for real-time closed-loop control. As shown in Table~\ref{tab:system_comparison}, \sysname addresses all three limitations.

\begin{table}[t]
  \centering
  \caption{{Comparison with State-of-the-Art}}\vspace{-4mm}
  \label{tab:system_comparison}
  \footnotesize 
  \setlength{\tabcolsep}{3pt} 
  \begin{adjustbox}{max width=\columnwidth}
  \begin{tabular}{l | c | c | c | c | c}
    \toprule
    \textbf{Works} & 
    \makecell{\textbf{Proactive}\\\textbf{Align?}} & 
    \makecell{\textbf{Radio}\\\textbf{Twin?}} & 
    \makecell{\textbf{Dynamic}\\\textbf{Blockage?}} & 
    \makecell{\textbf{NLoS}\\\textbf{Modeling?}} & 
    \makecell{\textbf{MU-}\\\textbf{MIMO?}} \\ 
    \midrule

    Agile-Link \cite{hassanieh2018fast} 
    & \xmark & \xmark & \xmark & \xmark & \xmark \\ \hline

    Two Beams \cite{jain2021two} 
    & \xmark & \xmark & \cmark & \cmark & \xmark \\ \hline

    M5 \cite{zhang2022m5} 
    & \cmark & \xmark & \xmark & \cmark & \cmark \\ \hline

    mmSV \cite{kamari2023mmsv} 
    & \cmark & \xmark & \xmark & \xmark & \cmark \\ \hline

    Commrad \cite{jain2024commrad} 
    & \cmark & \cmark & \cmark & \cmark & \xmark \\ \hline

    Lumos5G \cite{narayanan2020lumos5g} 
    & \cmark & \xmark & \xmark & \xmark & \xmark \\ \hline

    Habitus \cite{zhang2024habitus} 
    & \cmark & \xmark & \xmark & \xmark & \xmark \\ \hline

    MP2 \cite{xu2025roaming} 
    & \xmark & \xmark & \xmark & \xmark & \cmark \\ \hline
    
    RFCanvas \cite{chen2024rfcanvas} 
    & \xmark & \cmark & \cmark & \cmark & \xmark \\ \hline
    
    Radio Field \cite{zhao2023nerf2,zhang2026rf}
    & \xmark & \cmark & \xmark & \xmark & \xmark \\ \midrule

    \textbf{\sysname} 
    & \cmark & \cmark & \cmark & \cmark & \cmark \\
    \bottomrule
  \end{tabular}
  \end{adjustbox}\vspace{-6mm}
\end{table}
\textbf{XR mmWave Network System.} mmWave link quality prediction is critical due to its susceptibility to blockage. In XR, this is exacerbated by frequent body-induced shadowing and headset mobility. At the network layer, Lumos5G~\cite{narayanan2020lumos5g} uses dual-branch neural networks to predict throughput based on historical link data. Habitus~\cite{zhang2024habitus} improves this by incorporating full-body pose tracking to predict body-induced blockages. MP2~\cite{xu2025roaming} focuses on multi-path networking and roaming optimization in XR environments to mitigate signal drops. A parallel line of work uses viewport and head pose prediction~\cite{pan2025lookout,wang2024egonav,lai2017furion,qian2018flare} to optimize application-layer video delivery, deciding which video frames or tiles to transmit~\cite{guan2019pano,han2020vivo,jin2022where,liu2023cav3}. However, these systems operate at the network or application layer, treating the physical environment as a ``black box''. They react to link degradation rather than understanding the underlying physical reality. In contrast, \sysname replaces this black-box paradigm with a physics-guided real-time radio field that explicitly models geometry, scene dynamics, and multipath radio propagation, enabling proactive beam steering before outages occur.

\textbf{mmWave Beam Management.} Due to the fragility of mmWave links, standard protocols such as 802.11ad/ay~\cite{ghasempour2017ieee,m-cube} and 5G NR~\cite{giordani2018tutorial} rely on periodic exhaustive beam sweeps, incurring substantial overhead. To mitigate this, prediction-based methods extrapolate future beams from recent scanning history~\cite{hassanieh2018fast,zarifneshat2021learning,jain2021two,zhou2019robot}, but accuracy degrades rapidly under mobility. Motion-sensor-based approaches (e.g., IMU) attempt to track beam directions from device movement, yet they remain blind to environmental blockages---when a new obstacle occludes the path, these methods have no mechanism to detect or adapt. More recently, infrastructure-mounted sensors such as co-located radar~\cite{jain2024commrad} or cameras~\cite{xu2025computer, charan2021vision} provide environment-aware beam guidance, but are inherently limited to line-of-sight (LoS) scenarios; once the user moves into an NLoS region, the base station's sensors lose visibility. In contrast, \sysname leverages client-side egocentric perception to collaboratively maintain a real-time radio twin with the base station, enabling accurate beam prediction across both LoS and NLoS conditions even under dynamic blockages (Table~\ref{tab:system_comparison}).

\vspace{-2mm}\section{System Overview}
\label{s:system_overview}

\begin{figure}[t!]
    \centering
    \includegraphics[width=0.95\linewidth]{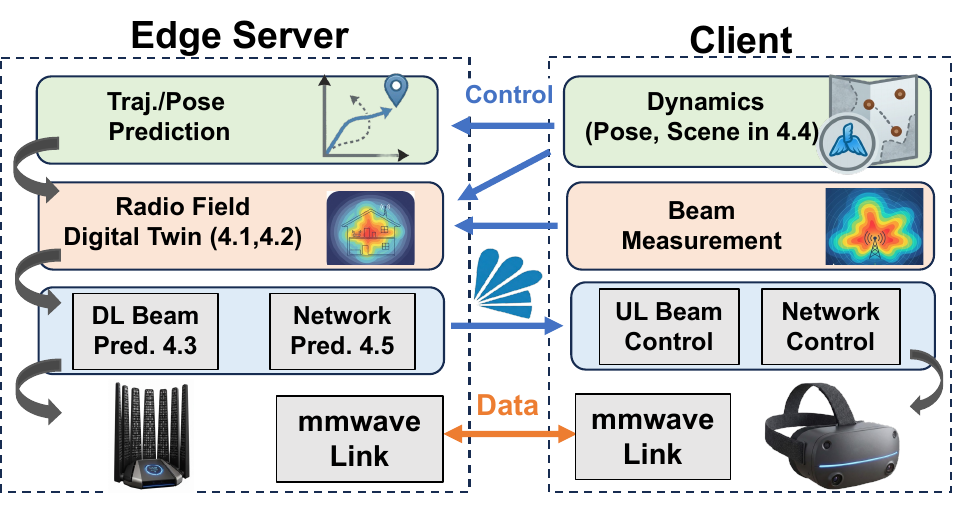}\vspace{-5mm}
    \caption{System Architecture. \textnormal{The Client streams environmental and user dynamics to the Edge Server. The Edge Server utilizes this data for trajectory prediction and maintaining a Radio Field Digital Twin, which in turn drives proactive beam prediction. }}\vspace{-5mm}
    \label{fig:system_framework}
\end{figure}

At the core of \sysname is a \textit{Radio Field Digital Twin}---a tightly synchronized, physics-guided model of the radio environment that resolves all valid propagation paths from the transmitter to any given receiver pose. Because each scheduling window's beams are decided during the preceding one (Fig.~\ref{fig:system_timeline}), \sysname must resolve the valid radio paths at future digital twin instances rather than at the present one. The validity of these paths is determined by three scene dynamics, which \sysname therefore tracks: the \textbf{UE pose}, the \textbf{transient blockers} (pose and 3D bounding-box volume), and the \textbf{registered movable objects}. To this end, \sysname adopts the client--server architecture of Fig.~\ref{fig:system_framework}.

The \textbf{XR Client} reports the dynamic metadata it observes: its own 6DoF pose, the 2.5D bounding boxes of any detected blockers, and a sparse spectrum scan (the optional scan of Fig.~\ref{fig:system_timeline}). Since the mmWave beamforming is derived entirely from the dynamic scene carried by these reports, they are transmitted over an omnidirectional 2.4\,GHz control channel rather than the mmWave link itself. The client simultaneously receives the directional mmWave XR video stream scheduled by the server.

The \textbf{Edge Server} fuses the reports from every client into a single dynamic scene, maintains the state of each tracked object, and synchronizes this state back to all UEs, so that every client shares a consistent view. It then queries the twin to derive the beam decisions: it renders the spatial spectrum at the predicted states, extracts the beam candidates, and computes the beamforming and rate-adaptation commands, which it dispatches to the Tx array and the UEs.

Three threads run in parallel on this architecture. Compute latencies quoted below are measured on our edge server (RTX~4090) for an indoor scene of $\sim$7$\times$9\,m (Appendix~\ref{sec:appendix_latency_analysis}).

\noindent$\blacksquare$ \textbf{Scene-Update Thread.} This asynchronous thread keeps the twin consistent with the physical scene and carries two latencies. The first is the one-time \textit{cold-start latency}: because the radio field is a per-environment twin, the client gathers visual and radio samples for the server to reconstruct the field once for each new environment (\S\ref{ss:radio_twin_const}). This scanning operates transparently during the XR headset's standard safety-boundary setup, requiring no extra user effort ($\sim$0.18\,s per scan, $\sim$30--60\,s of walking, and $\sim$1--2\,min for reconstruction). The second is the recurring \textit{scene-update latency} ($\sim$2.3\,ms of client-and-server processing). Whenever the UE pose shifts substantially or a registered object moves, the client triggers an update during an mmWave idle period, capturing a new visual frame and sparse spectrum. The server then fuses these updates with reported blocker bounding boxes, updates the global dynamic scene, and synchronizes it back to the UEs. Beam management never blocks on this thread, and simply reads the most recently updated scene.

\noindent$\blacksquare$ \textbf{Beam-Management Thread.} Within the current window, this thread decides the beams for the next one, at a cost we refer to as the \textit{beam-management latency} ($\sim$10.5\,ms). It reads the latest state maintained by the scene-update thread, namely the updated scene, the UE pose, and the blocker bounding boxes, and predicts each of the five 10\,ms frames within the coming 50\,ms window. For each window, it renders the spatial spectrum at every predicted frame instant for the two multiplexed UEs (2 UEs $\times$ 5 frames $\times$ 1.0\,ms, \S\ref{ss:radio_scene_registration}), back-traces each spectrum into Tx-side beam candidates ($0.3$\,ms, \S\ref{ss:beam_prediction}), and groups the MU-MIMO users and synthesizes the beamforming vectors ($0.2$\,ms, \S\ref{ss:traj_to_network_op}).

\noindent$\blacksquare$ \textbf{Motion-to-Photon Thread.} This thread renders, encodes, transmits, and displays the XR content itself, at a cost we refer to as the \textit{motion-to-photon (MTP) latency} ($\sim$29\,ms).

\begin{figure*}[t!]
    \centering
    \includegraphics[width=0.9\linewidth]{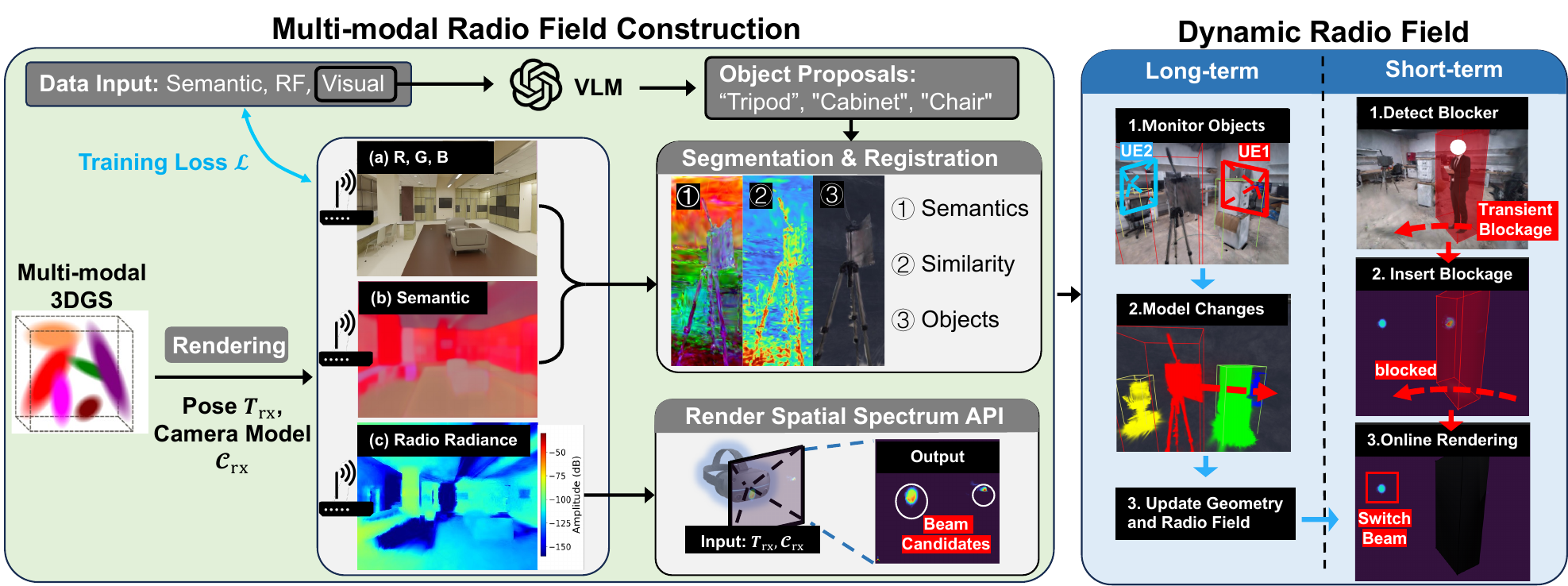}\vspace{-4mm}
    \caption{Multi-modal Radio Field and Dynamic Updates. \textnormal{The multi-modal 3DGS primitive encapsulates geometry along with visual, radio, and semantic attributes. The framework can simultaneously render (a) realistic RGB views, (b) semantic segmentation masks from the same underlying representation, and (c) radio spatial spectrum heatmaps. The radio spectrum in this figure are Sionna-RT simulated.}}\vspace{-4mm}
    \label{fig:radio_twin_construction}
\end{figure*}

\vspace{-2mm}\section{System Design} \label{s:system_design}

\vspace{-0.5mm}\subsection{Multi-modal Radio Field Construction}
\label{ss:radio_twin_const}
\vspace{-0.5mm}
Following~\cite{zhang2026rf,yang2026gsrf}, a neural radio radiance field models the radio environment as a set of $N$ radiance primitives $\mathcal{G} = \{g_i\}_{i=1}^{N}$ distributed throughout 3D space. Each primitive $g_i$ is a 3D Gaussian parameterized by its center $\mu_i \in \mathbb{R}^3$, covariance $\Sigma_i$ (encoding spatial extent and orientation), opacity $\alpha_i$, and a directional radio radiance function modeled via spherical harmonics. Given a receiver with a 6DoF pose $\mathbf{T}_{rx}$ representing 3D rigid body transformations, composed of a translation vector (position) and a rotation matrix (orientation), and a specific camera model $\mathcal{C}_{rx}$, the system queries the field. Essentially, $\mathcal{C}_{rx}$ defines the receiver's field of view and angular resolution as a discrete set of queried pixels $\mathcal{P} = \{p_m\}_{m=1}^M$. Each pixel $p_m \in \mathcal{P}$ corresponds to a specific observation ray mapped to an angle of arrival (AoA) denoted as $(\theta_{rx}^{(m)}, \phi_{rx}^{(m)})$. Across these explicit ray directions, the radio spatial spectrum (i.e., the angular power profile over all AoAs) is synthesized alongside other modalities via a general volume rendering function $\mathcal{R}_{\mathcal{G}, RF}(\mathbf{T}_{rx}, \mathcal{C}_{rx})$~\cite{mildenhall2021nerf}.

However, radio-only radiance fields~\cite{zhao2023nerf2, yang2026gsrf} train Gaussian primitives using RF measurements alone. Because wireless signals have inherently coarse angular and spatial resolution, RF-only supervision forces the model to learn both the scene geometry (i.e., the placement of $\mu$) and radio properties simultaneously from sparse data. This leads to poor data efficiency, slow convergence, and highly ambiguous geometry where primitives float freely rather than anchoring to physical surfaces. Such geometric degradation not only reduces spectrum synthesis accuracy but also breaks downstream tasks that require precise surface localization, such as ray back-tracing for AoD estimation (\S\ref{ss:beam_prediction}).

To address this, we extend the above radio radiance field into a \textit{multi-modal} 3DGS representation that jointly encodes three modalities (RGB, radio, and semantic) within a unified set of Gaussian primitives $\mathcal{G}$ (Fig.~\ref{fig:radio_twin_construction}). The key idea is that all three modalities \textit{share the same geometric backbone}: each Gaussian retains its center $\mu$, covariance $\Sigma$, and opacity $\alpha$, but is augmented with additional feature channels for each modality. Concretely, we extend each primitive from the original radio-only parameterization to a multi-modal form:
\begin{equation}\small\label{eq:multi_modal_primitive}
    g_i = \bigl\{\underbrace{\mu_i, \Sigma_i, \alpha_i}_{\text{shared geometry}},\; \underbrace{\mathbf{f}_{\text{RGB},i}(\mathbf{d}),\; \mathbf{f}_{\text{RF},i}(\mathbf{d}),\; \mathbf{f}_{s,i}}_{\text{multi-modal features}}\ \ \bigr\}
\end{equation}
where $\mathbf{f}_{\text{RGB},i}$ and $\mathbf{f}_{\text{RF},i}$ are view-dependent color and radio radiance explicitly conditioned on the viewing direction $\mathbf{d}$. The feature $\mathbf{f}_{s,i}$ ($s$ for semantic) is a view-independent dense semantic embedding vector aligned with vision-language foundation models (e.g., CLIP~\cite{qiu2024feature}) via contrastive learning, so that Gaussians on the same object learn similar semantic embeddings and are segmented into object clusters by the corresponding text query, as shown in Fig.~\ref{fig:radio_twin_construction}. Beyond anchoring segmentation, the semantic information also tells \sysname which objects are actually movable, so it can concentrate each dynamic update on the movable object's clusters instead of the whole field.

We further extend the classic radio or RGB rendering to a multi-modal pipeline. As shown in Fig.~\ref{fig:radio_twin_construction} (left), for a given receiver pose $\mathbf{T}_{rx}$ and camera model $\mathcal{C}_{rx}$, the rendered 2D map $\mathcal{P}_c$ for any target modality channel $c \in \{\text{RGB}, \text{RF}, s, \text{Depth}\}$~\footnote{Depth is inferred from the RGB-trained 3DGS~\cite{chen2024pgsr}.} is obtained via:
\begin{equation}\small\label{eq:multi_modal_rendering}
    \mathcal{P}_c = \mathcal{R}_{\mathcal{G}, c}(\mathbf{T}_{rx}, \mathcal{C}_{rx}) = \sum_{i=1}^{N} \mathbf{f}_{c,i} \, \alpha_i \, T_i, \quad T_i = \prod_{j=1}^{i-1}(1-\alpha_j)
\end{equation}
This rendering can produce four distinct outputs, each serving a critical system role:
\begin{enumerate}[leftmargin=*]
    \item \textit{RGB Image ($\hat{\mathbf{I}}=\mathcal{P}_{\text{RGB}}$):} Photorealistic views that enforce multi-view photometric consistency and force the shared Gaussian geometry $(\mu_i, \Sigma_i)$ to converge onto true physical surfaces---providing the accurate geometric foundation that all other modalities rely on.
    \item \textit{Radio Spatial Spectrum ($\hat{\mathbf{R}} = \mathcal{P}_{\text{RF}}$):} Rendered at any Rx configuration, capturing multipath signal distribution and intensity across all AoAs~\cite{zhang2026rf} for beam prediction. Because the primitives are geometrically anchored by RGB supervision, the synthesized spectrum benefits from accurate surface localization.
    \item \textit{Semantic Map ($\hat{\mathbf{S}}=\mathcal{P}_{s}$):} Language-aligned feature maps for object-level scene segmentation. As shown in Fig.~\ref{fig:radio_twin_construction}, by clustering Gaussians with similar $\mathbf{f}_{s}$ embeddings, \sysname decomposes the scene into discrete object groups and classifies them as static background or movable entities (see details in Appendix~\ref{segmentation}). This registration is the foundation for efficient dynamic updates (\S\ref{ss:dynamic_radio_field}).
    \item \textit{Depth Map ($\hat{\mathbf{D}}=\mathcal{P}_{\text{Depth}}$):} To accelerate downstream beam tracing (\S\ref{ss:beam_prediction}), we simultaneously render the expected depth map during the forward pass, where the depth feature $\mathbf{f}_{\text{Depth},i}$ of each Gaussian $g_i$ is the projected distance $d_i$ between the Rx and the Gaussian center.
\end{enumerate}

\begin{figure}[t!]
    \centering
    \includegraphics[width=0.95\linewidth]{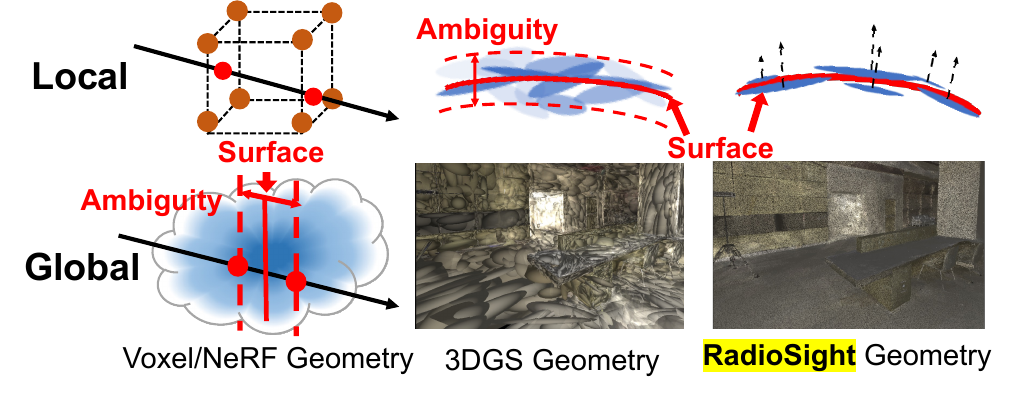}\vspace{-5mm}
    \caption{Geometry. \textnormal{Compared to previous radiance field methods with large geometry ambiguity, \sysname uses geometry regularization to force the Gaussians to deform into surfel-like shapes and align closely with the physical surface.}}\vspace{-4mm}
    \label{fig:geometry_demo}
\end{figure}

\vspace{-2mm}\subsection{Field Training and Rendering}
\label{ss:radio_scene_registration}

With the multi-modal primitives and differentiable renderer defined above, training the radio field reduces to designing appropriate supervision for each modality channel. Because the entire rendering pipeline is fully differentiable, we can compare the rendered outputs against their respective ground-truth measurements ($\mathbf{I}, \mathbf{R}, \mathbf{S}$) and back-propagate the gradients to update the geometry and the per-modality features. The three targets are defined on one shared pixel grid: we fix the visual pinhole camera model as $\mathcal{C}_{rx}$, extract the semantic feature map on its image plane, and project the mmWave measurements onto the same grid. Each rendered modality is therefore compared against a target sampled along identical ray directions. We detach the computation graph per modality, so the geometry ($\mu,\Sigma,\alpha$) receives gradients only from the visual loss, $\mathbf{f}_{RF}$ only from the radio loss, and $\mathbf{f}_s$ only from the semantic loss. We define the multi-modal training objective $\mathcal{L}$, the total loss summed over all pixels, as:
\begin{equation}\small\label{eq:total_loss}
    \mathcal{L} = \lambda_{RGB} \mathcal{L}_{RGB} + \lambda_{RF} \mathcal{L}_{RF} + \lambda_{sem} \mathcal{L}_{sem} + \lambda_{reg} \mathcal{L}_{reg}
\end{equation}
where each term supervises a specific aspect of the field:
\begin{enumerate}[leftmargin=*]
\item\textit{$\mathcal{L}_{RGB}$} compares rendered RGB images against captured photographs to establish photometric constraints that anchor the Gaussian geometry to real surfaces:
\begin{equation}\small
    \mathcal{L}_{RGB} = (1-\lambda)\|\hat{\mathbf{I}} - \mathbf{I}\|_1 + \lambda (1 - \text{SSIM}(\hat{\mathbf{I}}, \mathbf{I}))
\end{equation}
\item\textit{$\mathcal{L}_{RF}$} minimizes the $L_1$ error between the predicted radio spatial spectrum $\hat{\mathbf{R}}$ and the ground-truth mmWave sounding measurements $\mathbf{R}$, following RF-3DGS~\cite{zhang2026rf}:
\begin{equation}\small
    \mathcal{L}_{RF} = \|\hat{\mathbf{R}} - \mathbf{R}\|_1
\end{equation}
\item\textit{$\mathcal{L}_{sem}$} applies a contrastive loss to align the rendered semantic map $\hat{\mathbf{S}}$ with the pseudo-ground-truth embeddings $\mathbf{S}$. As shown in Fig.~\ref{fig:radio_twin_construction}, after each Gaussian learns its corresponding semantic feature through this alignment, we utilize the vision-language foundation model (VLM)~\cite{qiu2024feature} to extract the semantic embeddings of all movable object keywords identified from the visual input. By computing the feature similarity against these keyword embeddings, we cluster the Gaussians belonging to each movable object together. The loss function enabling this language-aligned segmentation is formulated as:
\begin{equation}\small
\mathcal{L}_{sem} = 1 - \frac{\hat{\mathbf{S}} \cdot \mathbf{S}}{||\hat{\mathbf{S}}||_2 ||\mathbf{S}||_2}
\end{equation}
\item\textit{$\mathcal{L}_{reg}$} adopts the geometric regularization from PGSR~\cite{chen2024pgsr}---enforcing properties such as minimizing the shortest axis scale---to further eliminate structural ambiguities and force the Gaussians to approximate thin, surfel-like shapes that tightly align with actual physical surfaces (see Appendix~\ref{sec::gaussian_regul}). As shown in Fig.~\ref{fig:geometry_demo}, without this regularization, primitives degenerate into volumetric blobs that break downstream ray back-tracing.
\end{enumerate}

\textbf{Real-Time Rendering and Inference.} Once trained, the radio field acts as a continuously queryable engine for network optimization. Because the underlying 3DGS representation is explicit, rendering is inherently fast and parallelizable. During RF inference, \sysname only needs to render the radio channel $\hat{\mathbf{R}}$ and the depth map $\hat{\mathbf{D}}$, skipping RGB and semantic computation entirely, which significantly reduces the per-query cost. Moreover, because physical-layer beamforming does not require pixel-perfect angular resolution, \sysname dynamically lowers the rendering resolution to match the array's beamwidth. Together, these optimizations cut the cost of rendering one spatial spectrum to $\sim$1.0\,ms, so serving one UE over a window costs $5 \times 1.0$\,ms for its five predicted frames (\S\ref{sec:beam_prediction_accuracy}).

\begin{figure}[t!]
    \centering
    \includegraphics[width=0.9\linewidth]{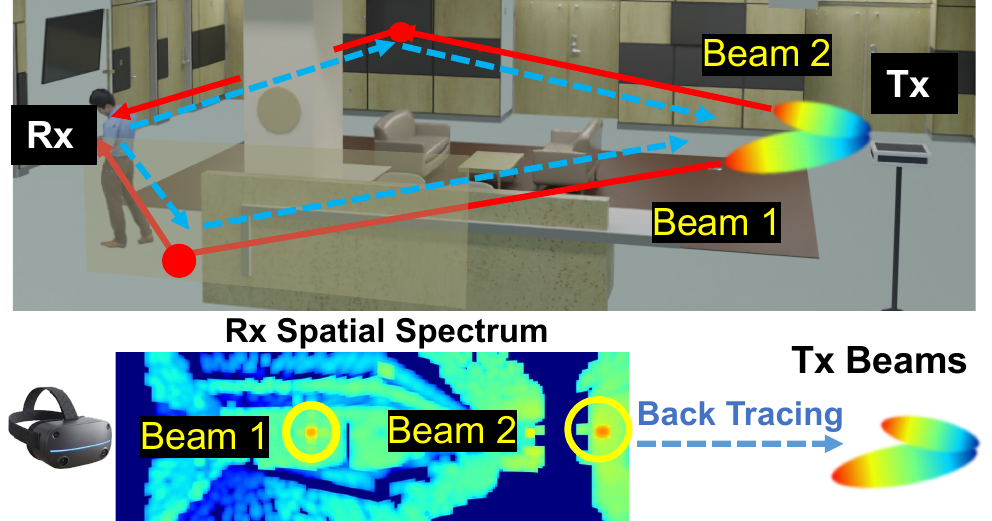}\vspace{-4mm}
    \caption{Predictive Beam Tracking. \textnormal{The server synthesizes the Rx spatial spectrum from the client's predicted viewpoint. By identifying dominant peaks in this spectrum, the system extracts the Top-$k$ AoAs and path gains. Finally, by back-tracing these paths through the scene, the optimal beam directions are derived. }}\vspace{-4mm}
    \label{fig:beam_prediction}
\end{figure}

\begin{figure}[t!]
    \centering
    \includegraphics[width=0.85\linewidth]{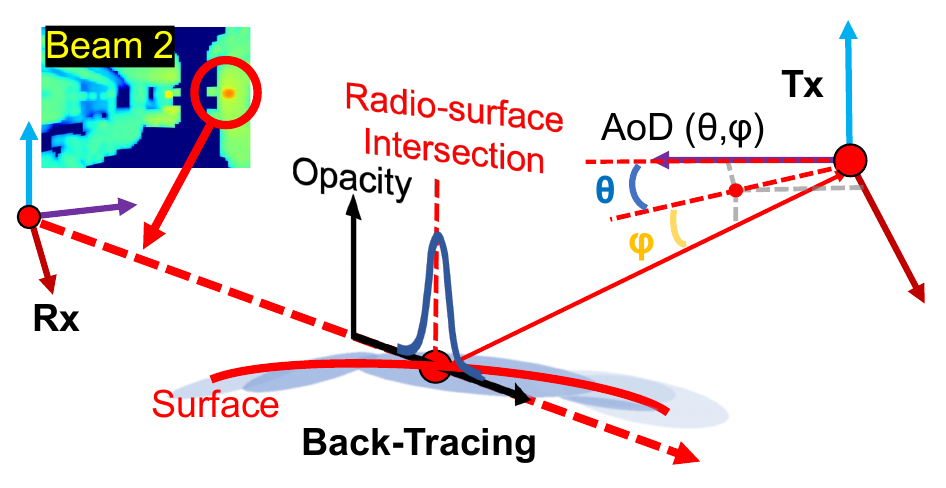}\vspace{-6mm}
    \caption{Back-tracing. \textnormal{For each detected signal peak, \sysname estimates the radio-surface intersection via accumulated opacity along the ray, subsequently tracing this point back to the Tx to determine the optimal AoD.}}\vspace{-6mm}
    \label{fig:back-tracing}
\end{figure}

\vspace{-2mm}\subsection{Backward Beam-Tracing}
\label{ss:beam_prediction}

The radio field constructed above renders the spatial spectrum from the receiver's perspective. While this provides rich multipath information, it alone cannot determine how to guide the \textit{transmitter's} phased array for downlink beamforming, since the Tx-side AoD is not directly observable from the Rx spectrum. To bridge this gap, \sysname introduces a backward beam-tracing pipeline (Fig.~\ref{fig:beam_prediction}) that translates Rx-side AoA peaks into Tx-side AoDs, enabling proactive beam scheduling without exhaustive channel-sounding scans.
The pipeline operates in three steps (Fig.~\ref{fig:back-tracing}):

\textbf{(1)~Spectrum query and peak detection.} Given a predicted Rx pose $\mathbf{T}_{rx}$ and its associated camera model $\mathcal{C}_{rx}$, the server renders the spatial spectrum $\hat{\mathbf{R}} = \mathcal{R}_{\mathcal{G}, \text{RF}}(\mathbf{T}_{rx}, \mathcal{C}_{rx})$ and the corresponding depth map $\hat{\mathbf{D}} = \mathcal{R}_{\mathcal{G}, \text{Depth}}(\mathbf{T}_{rx}, \mathcal{C}_{rx})$ in a single pass. It identifies the Top-$K$ signal peaks by locating local maxima in $\hat{\mathbf{R}}$. For the $k$-th peak, we explicitly define its angle of arrival as $\text{AoA}_k = (\theta_{rx}^{(k)}, \phi_{rx}^{(k)})$ and its corresponding received signal strength as $\text{RSSI}_k = \hat{\mathbf{R}}(\text{AoA}_k)$.

\textbf{(2)~Constant-time back-tracing.} Assuming single-bounce propagation~\cite{jain2024commrad} (our experiments suggest that single-bounce reflections generally dominate mmWave NLoS propagation in typical indoor environments; see Appendix~\ref{sec:single_bounce_energy}), \sysname maps these Rx arrival directions back to the physical environment. Because the depth map $\hat{\mathbf{D}}$ is rendered simultaneously with the spatial spectrum, determining the radio-surface intersection requires no iterative ray marching. Let $\mathbf{d}_k$ be the unit direction vector corresponding to $\text{AoA}_k$, and let $\mathbf{p}_{rx}$ be the translation component of the pose $\mathbf{T}_{rx}$. The intersection point, $\mathbf{x}_{int}^{(k)}$, is computed directly in $O(1)$ time:
\begin{equation}\small
    \mathbf{x}_{int}^{(k)} = \mathbf{p}_{rx} + \hat{\mathbf{D}}(\text{AoA}_k) \mathbf{d}_k
\end{equation}
Because the geometric regularization $\mathcal{L}_{reg}$ forces each Gaussian to tightly align with the true physical surface~\cite{chen2024pgsr}, $\mathbf{x}_{int}^{(k)}$ accurately reflects the actual reflection point.

\textbf{(3)~AoD derivation.} By tracing from the intersection point back to the transmitter location $\mathbf{p}_{tx}$, the departure vector is derived as $\mathbf{v}_{tx}^{(k)} = \mathbf{p}_{tx} - \mathbf{x}_{int}^{(k)}$. Converting this vector into the transmitter's local coordinate system yields the optimal departure angle, $\text{AoD}_k = (\theta_{tx}^{(k)}, \phi_{tx}^{(k)})$. The beam tracking thus returns the complete set of beam candidates:
\begin{equation}\small\label{eq:query_rf}
    \{(\text{AoA}_k, \text{AoD}_k, \text{RSSI}_k)\}_{k=1}^K = \mathcal{Q}(\mathbf{T}_{rx}, \mathcal{C}_{rx}, \mathbf{p}_{tx} \mid \mathcal{G})
\end{equation}
where $\mathcal{Q}$ denotes the comprehensive back-tracing query function over the multi-modal field $\mathcal{G}$, enabling the edge server to proactively configure its Tx array before the user arrives.

\vspace{-2mm}\subsection{Collaborative Dynamic Radio Field} \label{ss:dynamic_radio_field}

\begin{figure}[t!]
    \centering
    \includegraphics[width=0.95\linewidth]{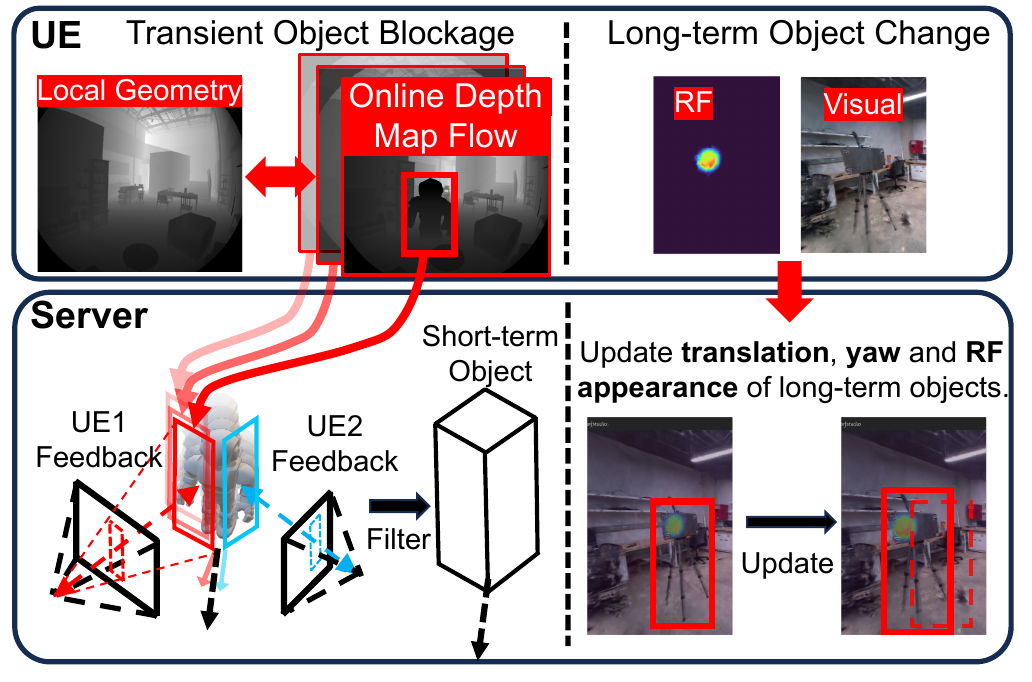}\vspace{-4mm}
    \caption{Dynamic RadioField Update. \textnormal{The server-side maintains per-object Gaussian clusters with 3D bounding boxes, enabling independent tracking and dynamic updates for blockage detection.}}\vspace{-6mm}
    \label{fig:object_registration}
\end{figure}

An original radio field is trained from spatial RF measurements. The straightforward way to keep it up-to-date when the environment changes is to re-collect measurements and retrain the model from scratch. This is clearly impractical for a real-time system. Our key insight is that the multi-modal XR systems already provide a rich visual--geometric understanding of the scene, and this understanding can be repurposed to drive radio field updates \textit{without} additional RF measurements (Fig.~\ref{fig:object_registration}). Concretely, \sysname decouples dynamic changes into two temporal regimes and handles each with an appropriately lightweight mechanism:

\noindent$\blacksquare$ \textbf{Short-term Transient Blockage} (e.g., a walking person): such fast movers only \textit{occlude} existing paths without introducing stable NLoS paths, so \sysname only tracks each blocker on the server as a moving 3D bounding box. A lightweight client--server loop maintains that box:
\textbf{(1)~Client-side detection.} Each UE renders a depth map from its server-synchronized local geometry and compares it against the online depth-map flow from its onboard sensors. It extracts the unmatched region as a compact 2.5D bounding-box proposal and reports it, with its periodic 6DoF pose, over the 2.4\,GHz control channel (upper-left of Fig.~\ref{fig:object_registration}).
\textbf{(2)~Server-side tracking.} A tracking-by-detection workflow~\cite{weng2020ab3dmot} associates the proposals from all UEs across frames. The server fuses the 2.5D bounding-box proposals that belong to transient blockers into a 3D bounding box, and predicts its motion with an Extended Kalman Filter (EKF). The blocker is now a tracked box in the dynamic radio field; detection through fusion takes $\sim$2.3\,ms of processing, and $\sim$3.8\,ms end-to-end including the uplink hop.

\noindent$\blacksquare$ \textbf{Long-term Object Change} (e.g., a relocated tripod): the long-term object is a \textit{movable object} semantically segmented during multi-modal reconstruction (\S\ref{ss:radio_twin_const}). Its displacement also changes the NLoS paths it introduced, and based on the segmentation, \sysname can rapidly edit its pose rather than retraining the field:
\textbf{(1)~Client-side detection and optional scan.} The server also synchronizes each movable object's 3D bounding box to the UE's local geometry. During the UE's online depth comparison, it can likewise detect such an object's change as a divergence from its expected 3D geometry, which raises an mmWave scan request---the same \textit{sparse optional scan} triggered when the UE's pose shifts substantially (Fig.~\ref{fig:system_timeline}). The request is served when the mmWave channel is idle or has abundant resources, and the UE then uploads an aligned visual frame and radio spectrum (upper-right of Fig.~\ref{fig:object_registration}). The server runs the same multi-object tracking workflow~\cite{weng2020ab3dmot,li2026dynagslam} to classify the change as a registered-object movement or a new transient blocker based on global information.
\textbf{(2)~Real-time field editing.} For a confirmed movement, \sysname rigidly transforms the object's Gaussian cluster to its new pose and refits its radio appearance from the uploaded samples, leaving the rest of the field untouched.

\noindent$\blacksquare$ \textbf{Serving the Updated Field.} The real-time radio field closes two loops. \textit{First}, the server pushes the refreshed box geometry to every UE, synchronizing the local geometry each client compares against. \textit{Second}, the updated field immediately feeds the next beam management horizon~(\S\ref{ss:traj_to_network_op}): each tracked box is advanced to its predicted pose, so that (i)~every movable object renders at its latest location, exposing the valid LoS and NLoS paths, and (ii)~each transient blocker enters as an opaque 3D box ($\alpha=1$, $\mathbf{f}_{RF}=0$) that attenuates any path through it, masking the LoS and the reflector$\to$Rx segment of each NLoS path in the render. Because 3DGS rendering is receiver-centric, the Tx$\to$reflector segment is invisible to it; a lightweight ray--box test on each back-traced candidate (\S\ref{ss:beam_prediction}) discards those whose incident segment is blocked.

While prior work such as RFCanvas~\cite{chen2024rfcanvas} also attempts to maintain temporal consistency by fusing vision and RF data, it relies on external cameras and NeRF-like updates that require seconds of processing---fundamentally too slow for real-time beam management. In contrast, \sysname's update architecture is designed to operate within the latency budget of live XR streaming. From detecting a blocker, through its appearance in the dynamic field, to the farthest beamforming scheduled against it, the chain spans the update latency plus a 100\,ms prediction reach. The prediction error in \S\ref{sec:beam_prediction_accuracy} shows the EKF mobility model stays accurate at this horizon.

\vspace{-2mm}\subsection{Proactive Operation and Multi-User AP} \label{ss:traj_to_network_op}
\begin{figure}[t]
    \centering
    \includegraphics[width=0.95\linewidth]{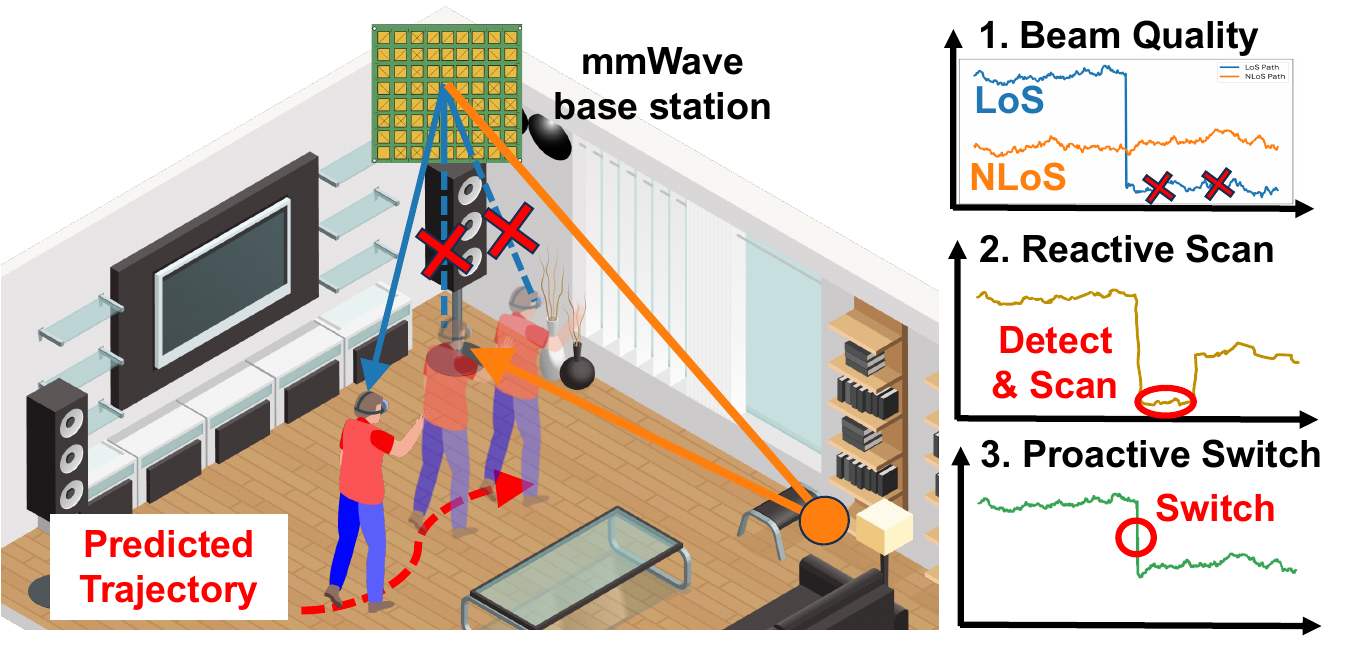}\vspace{-5mm}
    \caption{Predictive Trajectory to Proactive Control. \textnormal{The system translates predicted user trajectories into proactive beam scheduling and adaptive bitrate control for low-latency XR delivery.}}\vspace{-6mm}
    \label{fig:scenario_demo}
\end{figure}

With a continuously synchronized radio twin, \sysname can go beyond passive channel estimation: by predicting where users and dynamic objects \textit{will be}, the system queries the twin at future poses to obtain optimal beam configurations \textit{before} they are needed (Fig.~\ref{fig:scenario_demo}). This is fundamentally different from prior black-box approaches~\cite{narayanan2020lumos5g,zhang2024habitus} that predict future link quality from historical signal sequences---such methods can forecast \textit{that} a link will degrade, but cannot identify the best alternative path without exhaustive scanning. In contrast, because the radio twin encodes the full spatial spectrum, a single query at a predicted pose directly reveals all viable LoS and NLoS beam candidates along with their AoD, AoA, and RSSI. As illustrated in Fig.~\ref{fig:scenario_demo}, even while the current LoS path is still strong, the twin detects that the user's future trajectory will be occluded and proactively steers the Tx toward an NLoS reflection path---executing a seamless beam switch before throughput degradation occurs.

\noindent$\blacksquare$ \textbf{Trajectory Prediction.} To realize this proactive pipeline, the server must predict the future state of \textit{both} users and dynamic objects, and then translate these predictions into beam decisions. For user trajectory, we adopt a kinematic motion model similar to M5~\cite{zhang2022m5}: the headset's SLAM service streams 6DoF poses to the edge server, which applies an EKF to extrapolate a sequence of future poses $\hat{\mathbf{T}}_{rx, t_j}$ over a configurable horizon (e.g., 50--100\,ms for immediate beam switching, 1--3\,s for multi-AP handover planning). For dynamic blockers detected in \S\ref{ss:dynamic_radio_field}, the server likewise propagates their EKF state vectors to predict future bounding-box.

\noindent$\blacksquare$ \textbf{Batch Spectrum Query.} Given these predicted trajectories, \sysname performs a \textit{batch query} to the radio twin. For each future time-step $t_j \in \{t_{i+1}, \dots, t_{i+n}\}$, the server updates the dynamic bounding boxes to their predicted positions, then renders the spatial spectrum $\hat{\mathbf{R}}_{u, t_j} = \mathcal{R}_{\mathcal{G}_{t_j}, \text{RF}}(\hat{\mathbf{T}}_{rx, u, t_j}, \mathcal{C}_{rx})$ at each UE $u$'s predicted pose $\hat{\mathbf{T}}_{rx, u, t_j}$ via the backward beam-tracing pipeline (\S\ref{ss:beam_prediction}). The output is a time-indexed candidate set for each UE $u$:
\begin{equation}\small
    \mathcal{S}_{u,t_j} = \{(\text{AoA}_k, \text{AoD}_k, \text{RSSI}_k)_{u,t_j}\}_{k=1}^K = \mathcal{Q}(\hat{\mathbf{T}}_{rx,u,t_j}, \mathcal{C}_{rx}, \mathbf{p}_{tx} \mid \mathcal{G}_{t_j})
\end{equation}
where $\mathcal{S}_{u,t_j}$ denotes the candidate set (avoiding notation conflict with the camera model $\mathcal{C}_{rx}$), and $\mathcal{G}_{t_j}$ represents the radio field geometry augmented with the predicted dynamic blockers at time $t_j$. Because all queries share the same static Gaussian geometry and differ only in the predicted poses and blocker positions, \sysname parallelizes the entire sequence using NeRFstudio-accelerated CUDA kernels, synthesizing the full batch for multiple UEs in real time.

\begin{figure}[t!]
    \centering
    \includegraphics[width=0.95\linewidth]{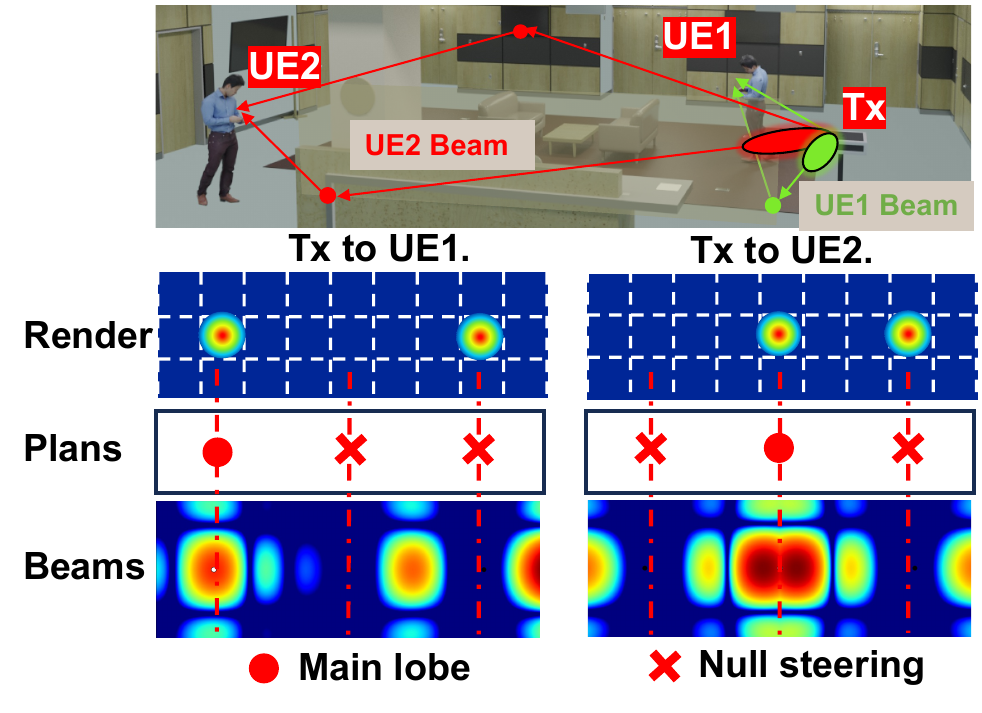}\vspace{-6mm}
    \caption{MU-MIMO Beam Management. \textnormal{The server renders per-UE spatial spectra from the radio field and jointly optimizes beam-steering vectors across multiple users to maximize the sum-rate while minimizing inter-user interference by null-steering.}}\vspace{-4mm}
    \label{fig:mu_mimo}
\end{figure}

\noindent$\blacksquare$ \textbf{MU-MIMO Beam Management and Scheduling.} Leveraging the batch query results, the edge server formulates a joint MU-MIMO scheduling and beamforming problem (Fig.~\ref{fig:mu_mimo}). For a group of UEs $\{u_1, u_2, \dots \}$ scheduled on the same time-frequency resource, the server possesses their predicted candidate sets $\mathcal{S}_{u,t_j}$ containing all viable propagation paths. The core challenge is to maximize sum-rate while minimizing inter-user interference. \sysname achieves this by \textit{overlaying} the rendered spatial spectra of all users in the group. As shown in the ``Render'' and ``Plans'' layers of Fig.~\ref{fig:mu_mimo}, the server identifies ``orthogonal'' path combinations where one user's signal peak (main-lobe) aligns with other users' low-energy regions. Based on these overlaid profiles, the server proactively selects the primary steering vector for each UE and calculates the necessary null-steering directions to suppress potential interference from other users' multipath components (MPCs). 

Finally, the server applies a codebook-based beam synthesis or the Linearly Constrained Minimum Variance (LCMV) algorithm to form these beam patterns (shown in the ``Beams'' layer). By anticipating user motion and blockage, \sysname can also trigger proactive \textit{Bitrate Adaptation}, lowering the video resolution for the UE just before its signal-to-interference-plus-noise ratio (SINR) is predicted to drop below the quality of experience (QoE) threshold. Technical details of the heuristic scheduling and interference nulling are in Appendix~\ref{sec:mu-mimo_beamforming}.

\noindent$\blacksquare$ \textbf{Multi-AP Handover.} For users traversing large-scale environments, \sysname extends its proactive foresight to access point (AP) coordination. As a UE moves toward a coverage boundary, the edge server dispatches the long-term predicted trajectory (\S\ref{ss:traj_to_network_op}) to all candidate APs in the vicinity. Leveraging their local radio twins, these APs parallel-query their respective spatial fields to evaluate exactly when and where the handover should occur. Instead of waiting for a signal drop to trigger a search, the servers proactively align their schedules: the target AP prepares its optimal beam configurations in advance, while the source AP maintains the link until the precise moment of transition. This collaborative prediction ensures zero-interruption handovers even during rapid movement or complex multi-path dynamics.

\vspace{-2mm}\subsection{Discussion}\label{sec:discussion}
\noindent$\blacksquare$ \textbf{Multi-User Visual Synergy and Blind-Spot Recovery.}
Although egocentric vision is inherently bounded by the wearer's optical field of view (FoV), this minimally impacts beam prediction since severe human-body attenuation blocks rear-arriving mmWave signals, largely aligning the effective \textit{radio FoV} with the visual FoV. However, dynamic occlusions outside this view still occur. To address this, \sysname acts as a collaborative sensing hub that mitigates individual blind spots. Specifically, it (1) fuses fragmented visual proposals from multiple XR users into a globally consistent scene representation, and (2) integrates infrastructure-side sensors (e.g., cameras~\cite{charan2021vision,zheng2025beamllm} or radar~\cite{jain2024commrad}) for persistent monitoring. Through this \textit{crowd-assisted sensing} paradigm, \sysname maintains robust environment awareness beyond any single device's FoV, ensuring reliable beam decisions under rapid motion and occlusions.

\noindent$\blacksquare$ \textbf{Cross-layer optimization.} 
Beyond physical-layer beam switching, \sysname's predictive spatial spectra enable proactive cross-layer enhancements. By forecasting future link capacity and SINR, the edge server empowers the application layer to shift from reactive bitrate adaptation to proactive throughput-aware pre-fetching~\cite{zhang2022m5}. For instance, if an imminent blockage is predicted, the XR streaming engine can preemptively buffer frames or adjust encoding parameters while link quality remains high. This deep integration between physical-layer channel dynamics and application-layer flow control ensures a resilient, zero-perceived-latency XR experience against sudden environmental fluctuations.

\noindent$\blacksquare$ \textbf{Physics-guided radio field.} 
Rather than functioning as a rigorous electromagnetic solver, \sysname implements a physics-guided radio field for practical deployment. First, while our 3DGS backbone provides a statistical geometric representation rather than a water-tight manifold, it is entirely sufficient for accurately modeling object boundaries, NLoS paths, and complex occlusion relationships. Second, \sysname intentionally predicts radio path strength without phase. This is a practical design choice: because our system relies on optimal single-beam transmission, it bypasses the need to compute coherent multi-path interference. A single directional beam provides sufficient capacity for high-demand XR content streaming while allowing the system to preserve structural simplicity and robustness.

\begin{figure}[t!]
    \centering
    \includegraphics[width=0.87\linewidth]{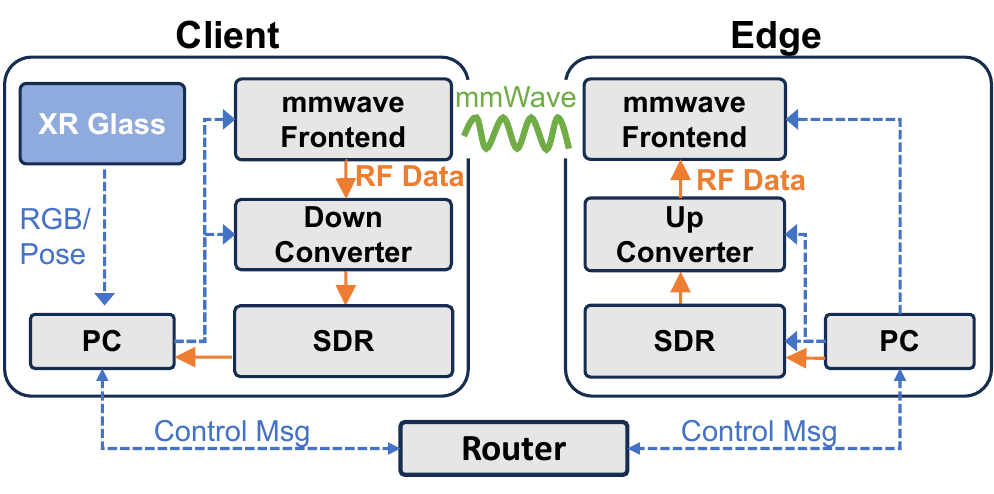}\vspace{-5mm}
    \caption{System Hardware Overview. \textnormal{The client side comprises XR glasses that stream RGB frames and 6DoF poses to a local PC, while an mmWave frontend with a down-converter and SDR captures RF data. The edge side mirrors this chain. A router interconnects both sides for control messaging and data exchange.}}\vspace{-6mm}
    \label{fig:system_hardware}
\end{figure}

\vspace{-2mm}\section{Implementation}
\label{sec:implementation}

\begin{figure}[t!]
      \centering
      \includegraphics[width=0.95\linewidth]{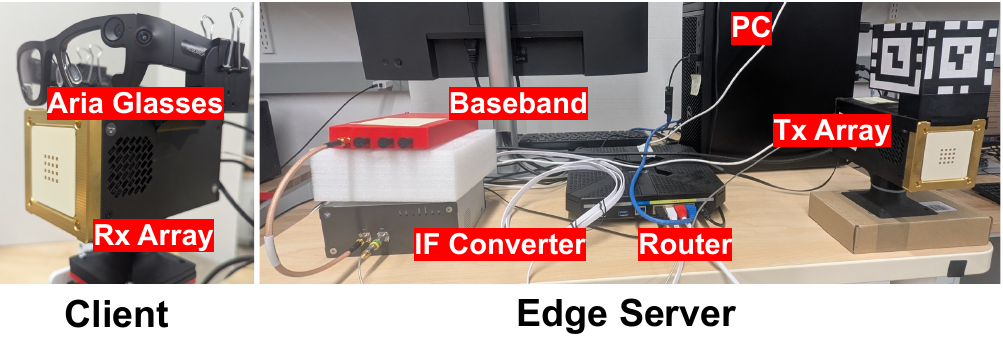}\vspace{-6mm}
      \caption{Hardware Prototype. \textnormal{The 28\,GHz mmWave testbed consists of an on-glass mmWave client and an mmWave edge node.}}\vspace{-6mm}
      \label{fig:hardware_prototype}
\end{figure}

We engineer an end-to-end prototype of \sysname by integrating a Radio Field engine with a real-world 28\,GHz mmWave testbed, as illustrated in the system architecture (Fig.~\ref{fig:system_hardware}) and the hardware deployment (Fig.~\ref{fig:hardware_prototype}).

\noindent$\blacksquare$ \textbf{Client-Side Architecture.} For the XR interface, we employ Meta's Project Aria glasses~\cite{meta2024aria}, which use Meta's SLAM service to stream RGB frames and 6DoF poses to a local client PC (Intel NUC11 i5). For the radio interface, the client acts as the physical Rx. We mount a TMYTEK BBox One5G mmWave $4 \times 4$ phased array~\cite{tmytek2024bbox} directly onto a headset rig to capture the spatial spectrum. The 28\,GHz RF signals are stepped down by a dedicated down-converter and digitized by an Ettus USRP B210 software-defined radio (SDR). To capture the mmWave channel characteristics with a low-power mobile CPU, we operate the SDR with a 10\,MHz instantaneous bandwidth. Shifting the intermediate frequency (IF) allows for synthesizing an effective 320\,MHz channel. The local PC acts as the client hub, processing both the RGB frames and the decoded RF samples into control messages which are routed to the edge server via the control channel.

\noindent$\blacksquare$ \textbf{Edge-Side Architecture.}
The edge node consists of another mmWave radio chain acting as the Tx with a PC (Intel\,Core\,i7) and a co-located server. The edge server (RTX 4090) runs the CUDA-accelerated render engine, which continuously updates the neural radio field from the client's control messages and implements proactive network operations. Once the optimal downlink beamforming is derived, the server dispatches control messages to its local baseband SDR and UE.
The generated baseband signal is then up-converted and fed to the Tx mmWave phased array. This $4 \times 4$ Tx array supports rapid beam steering, translating the Radio Field's proactive decisions into instantaneous physical radio transmissions. For beamforming, the testbed operates with a fast beamforming algorithm (see Appendix~\ref{app:codebook}).

\begin{figure*}[t!]
     \centering
     \begin{subfigure}[b]{0.44\linewidth}
         \centering
         \includegraphics[width=\linewidth]{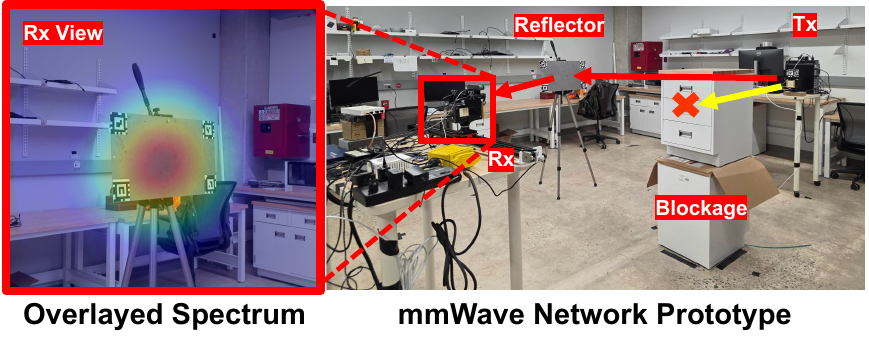}\vspace{-3mm}
         \caption{Radio Spectrum and Testbed Setup.}\vspace{-5mm}
     \end{subfigure}%
     \hfill
     \begin{subfigure}[b]{0.48\linewidth}
         \centering
         \includegraphics[width=\linewidth]{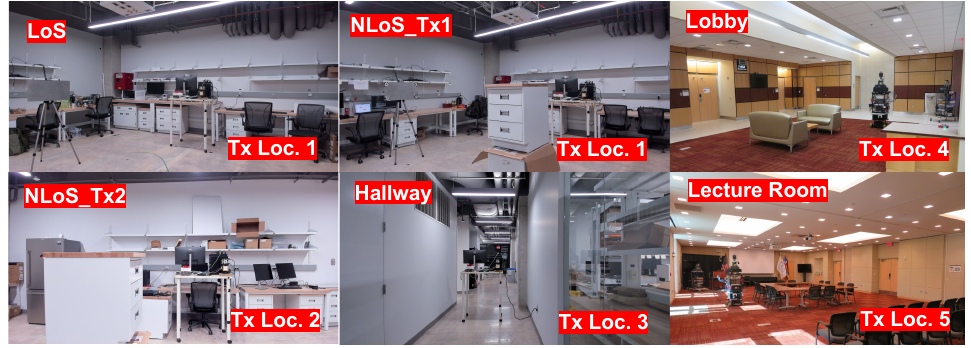}\vspace{-2mm}
         \caption{Experimental Scenes.}\vspace{-5mm}
     \end{subfigure}%
     \caption{Experimental Scenario. \textnormal{(a)~Left: Radio spectrum heatmap at the Rx; Right: Physical testbed with Tx/Rx, metal reflector, and blockage. (b)~The six experimental scenes spanning five Tx locations: LoS, NLoS (Tx1), NLoS (Tx2), Hallway, Lobby, and Lecture Room.}}\vspace{-5mm}
     \label{fig:experiment_scene}
\end{figure*} 

\begin{figure*}[t!]
     \centering
     \begin{subfigure}[b]{0.35\textwidth}
         \centering
         \includegraphics[width=\textwidth]{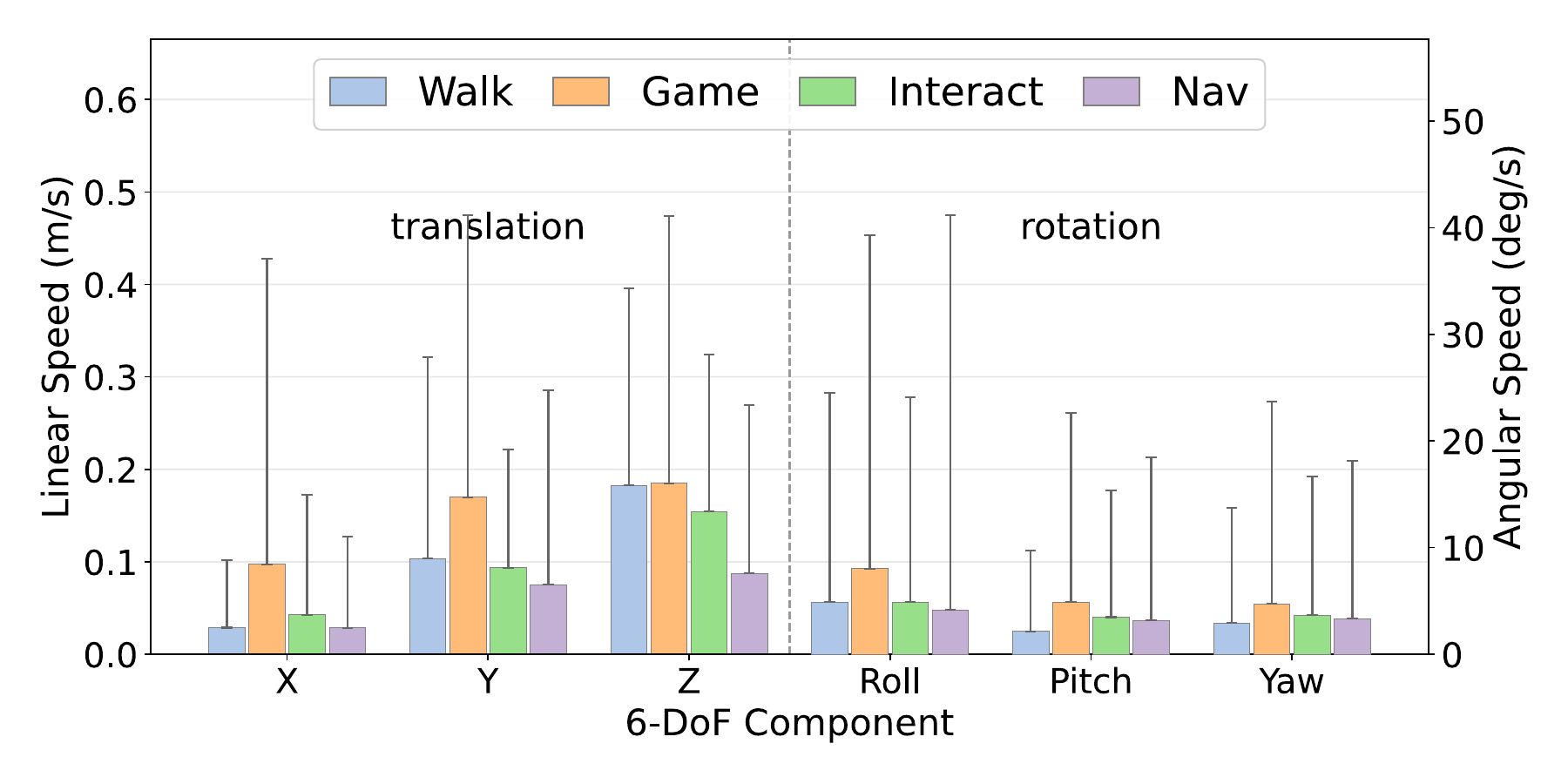}\vspace{-4mm}
         \caption{6-DoF Motion Statistics.}\vspace{-5mm}
     \end{subfigure}%
     \hfill
     \begin{subfigure}[b]{0.35\textwidth}
         \centering
         \includegraphics[width=\textwidth]{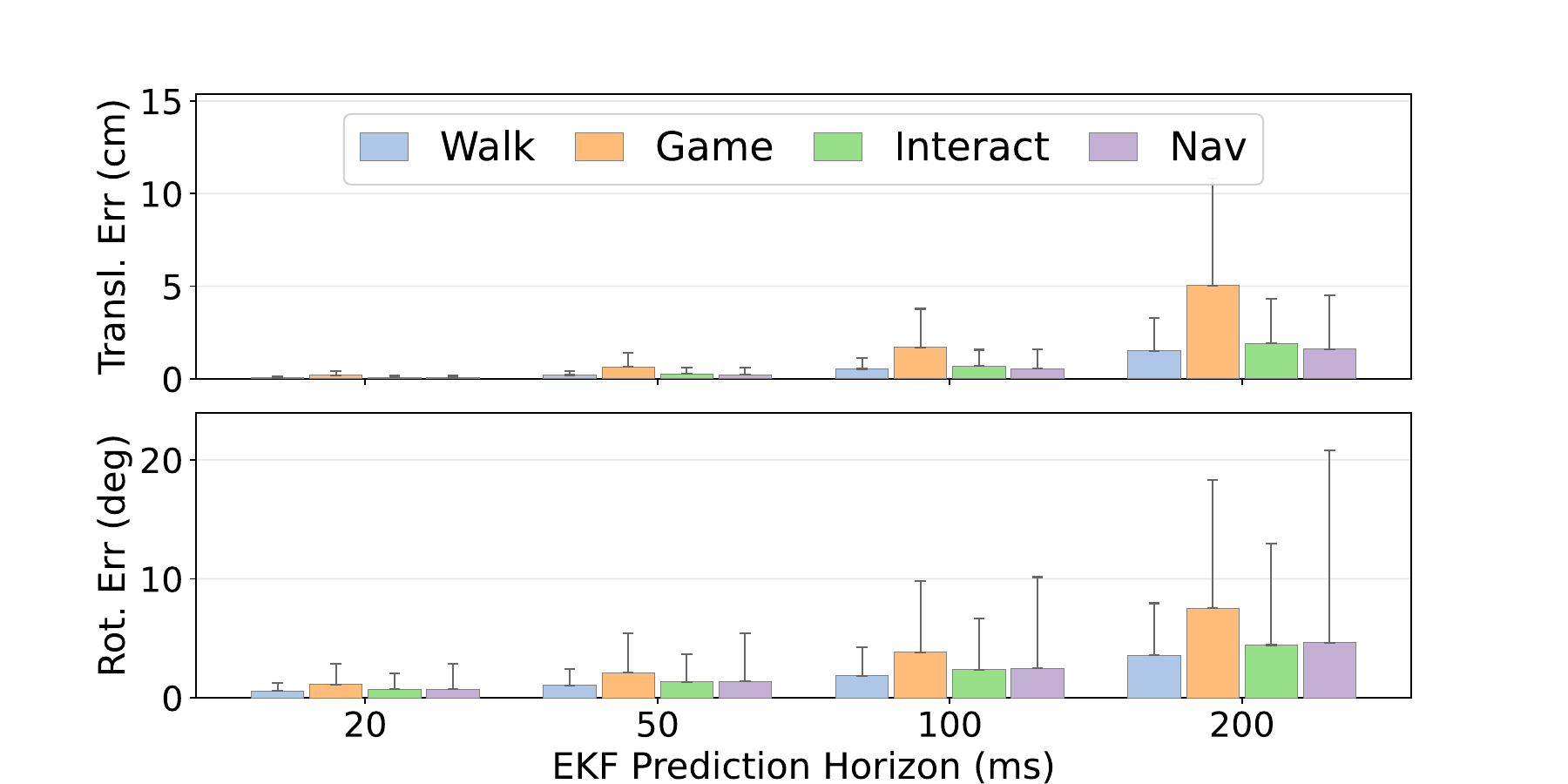}\vspace{-2mm}
         \caption{Pose-EKF Prediction Error.}\label{fig:pose_ekf_horizon}\vspace{-5mm}
     \end{subfigure}%
     \hfill
     \begin{subfigure}[b]{0.23\textwidth}
         \centering
         \includegraphics[width=\textwidth]{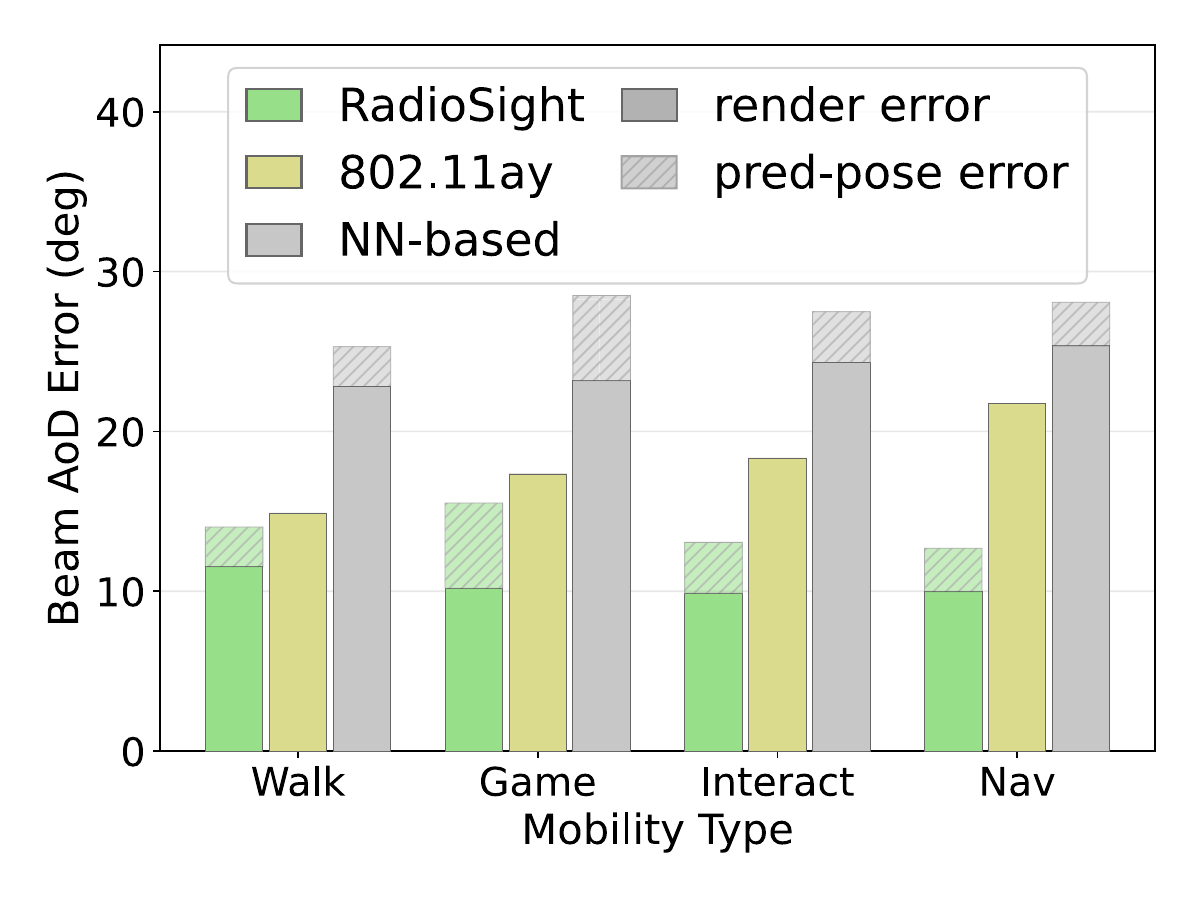}\vspace{-4mm}
         \caption{Beam Error by Mobility.}\label{fig:beam_mobility}\vspace{-5mm}
     \end{subfigure}%
     \caption{Beam Prediction under Mobility. \textnormal{(a)~Per-axis 6-DoF motion (median bar, p90 whisker) across four XR-mobility types. (b)~Pose-EKF prediction error versus horizon. (c)~Beam-AoD error by mobility type for baselines; RadioSight and NN-based decompose into a render error floor (solid) plus the pose-prediction penalty (hatched).}}\vspace{-5mm}
    \label{fig:beam_accuracy}
\end{figure*}

\noindent$\blacksquare$ \textbf{Experimental Scenarios and Datasets.}
As depicted in Fig.~\ref{fig:experiment_scene}, we deploy our testbed across indoor environments that differ significantly in geometry and material: an open laboratory under a direct-LoS layout, two NLoS layouts, and a long corridor that bends at a corner. 
We also collect the channel under mobility traces during typical XR tasks (e.g., walking, gaming, navigation, and interaction, shown in Fig.~\ref{fig:beam_accuracy}(a)).
We also leverage 28\,GHz measurements from a per-element phased-array channel sounder~\cite{caudill2021real}, whose $8\times8$ transmit and $8\times8$ receive arrays record the full per-element channel response, captured in a large lobby and a lecture room (Fig.~\ref{fig:experiment_scene}). Altogether, these measurements cover six distinct environments, spanning LoS, corridors, NLoS, and large open spaces.

\vspace{-2mm}\section{Evaluation}
\label{sec:evaluation}

\subsection{Experimental Setup}\label{sec:experimental_setup}

\noindent$\blacksquare$ \textbf{Ground Truth.} For each measurement location, the ground-truth beamforming configuration is determined by exhaustively sweeping the entire beamforming space to find the optimal beam. The ground-truth throughput is then evaluated based on this optimal beam.

\noindent$\blacksquare$ \textbf{Baselines.} We compare \sysname against 5 SoTA systems: (1) \textit{802.11ay Beam Scan}~\cite{ghasempour2017ieee}, which selects beams based on local scans for beam tracking and exhaustive sector sweeps for initialization and realignment; (2) \textit{End-to-End Neural Network} uses neural networks like LSTMs or RNNs on previous sequences (e.g., Lumos5G~\cite{narayanan2020lumos5g} and Habitus~\cite{zhang2024habitus}), where we adapt Lumos5G to convert the prediction of channel quality and throughput into the prediction of beam and throughput; (3) \textit{Vision-aided Beam Tracking}, which relies on router-side sensing to track UE dynamics and reflectors (e.g., CommRad~\cite{jain2024commrad}); to simulate CommRad's radar or vision system, we install a camera on the router, and when the UE leaves the router-camera field of view or visual tracking becomes unreliable, the baseline rescans a small set of beams around its last known beam (802.11ay-style local tracking); (4) \textit{NeRF-based Radiance Field: NeRF$^2$~\cite{zhao2023nerf2,tancik2023nerfstudio}}: As an ablation of the RF-only method, the density MLP and RF appearance MLP of NeRF$^2$ are trained only on the radio spectrum; (5) \textit{3DGS-based Radiance Field: WRF-GS~\cite{wen2025wrf}}, which we similarly initialize using SLAM point clouds.

\begin{figure}[t!]
    \centering
    \begin{subfigure}[b]{0.23\textwidth}
        \centering
        \includegraphics[width=\textwidth]{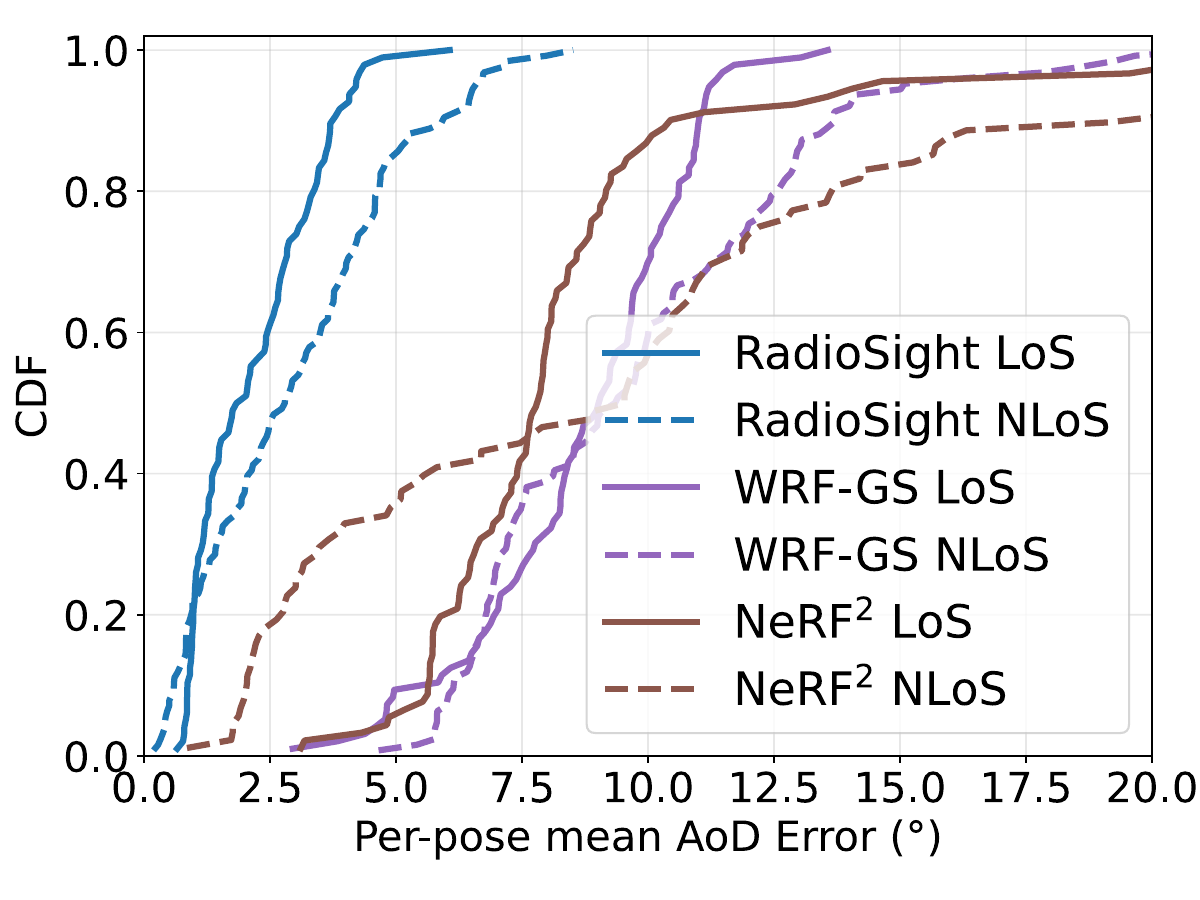}\vspace{-3mm}
        \caption{Region Beam Accuracy.}\vspace{-4mm}
        \label{fig:beam_accuracy_cdf}
    \end{subfigure}
    \hfill 
    \begin{subfigure}[b]{0.23\textwidth}
        \centering
        \includegraphics[width=\textwidth]{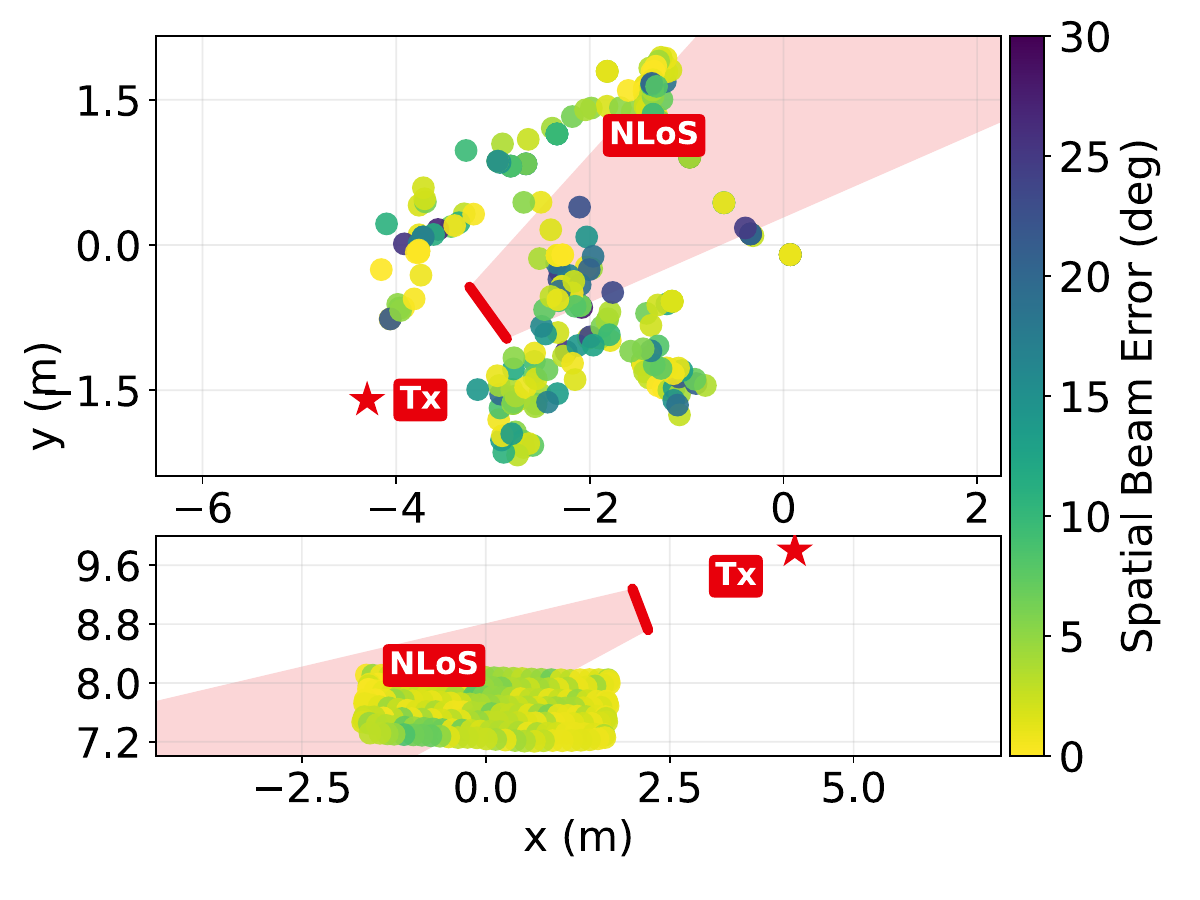}\vspace{-3mm}
        \caption{Spatial Distribution.}\vspace{-4mm}
        \label{fig:beam_accuracy_spatial}
    \end{subfigure}
    \caption{Static Beam Accuracy. \textnormal{(a)~\sysname outperforms NeRF$^2$ (ablation of RF-only), and WRF-GS (ablation of RF+simple RGB) in both LoS and NLoS regions. (b)~Spatial distribution of beam error in free trajectory (upper) and dense scan grid (lower).}}\vspace{-6mm}
\label{fig:overall_beam_accuracy}
\end{figure}

\vspace{-2mm}\subsection{Beam Prediction Accuracy}\label{sec:beam_prediction_accuracy} 
\noindent$\blacksquare$ \textbf{Static Beam Accuracy.} We first evaluate the beam prediction accuracy, where the Rx remains stationary at each location. We evaluate the AoD of transmitting beams. (Since the similarity of the rendered Rx-side spectrum has minimal impact on downstream system performance, the related analysis is deferred to Appendix~\ref{sec:spectrum_prediction_similarity}). As shown in Fig.~\ref{fig:overall_beam_accuracy}, \sysname attains a median beam-AoD error of 1.9$^\circ$ in the LoS region and 2.8$^\circ$ inside the shadowed NLoS region. The RF-supervised baselines are consistently worse and degrade further in the shadow: WRF-GS (RF+Simple RGB) yields 9.0$^\circ$ (LoS) and 9.4$^\circ$ (NLoS), while NeRF$^2$ (RF-only) yields 7.9$^\circ$ (LoS) and 11.9$^\circ$ (NLoS). \sysname's advantage stems from its visually trained geometry: our backward beam-tracing projects each reflection path from the accurate scene geometry to precisely locate the optimal transmitter AoD, which the RF-supervised fields cannot recover.

\noindent$\blacksquare$ \textbf{Pose Prediction Accuracy.} Similar to M5~\cite{zhang2022m5}, \sysname predicts the user's future 6-DoF rendering pose from a historical window to support proactive scheduling (Fig.~\ref{fig:pose_ekf_horizon}). We evaluate the EKF prediction error across horizons of 20--200\,ms. Benefiting from our low beam-management latency, \sysname natively requires a short prediction horizon: at the 20--50\,ms horizons the median translation error is only 0.1--0.6\,cm and the rotation error 0.6--2.1$^\circ$. Even at the longest required 100\,ms horizon, translation stays at 0.5--1.7\,cm and rotation at 1.9--3.8$^\circ$ across the four mobility types. Because our scheme lets us schedule at short horizons, these prediction errors remain well within the beamwidth tolerance.

\noindent$\blacksquare$ \textbf{Impact of Mobility.} We break beam prediction down across four real XR-mobility traces (walking, gaming, interaction, and navigation), whose per-axis 6-DoF motion statistics are shown in Fig.~\ref{fig:beam_accuracy}(a). Across all four types (Fig.~\ref{fig:beam_mobility}), \sysname holds a median beam-AoD error of 12.7--15.5$^\circ$, which decomposes into a 9.9--11.5$^\circ$ render error floor plus a 2.5--5.3$^\circ$ penalty of predicted future pose. The reactive 802.11ay scan ranges from 14.9$^\circ$ (walking) to 21.7$^\circ$ (navigation), and the NN-based predictor is worst at 22.8--25.3$^\circ$. \sysname therefore reduces beam error by up to $\sim$50\% relative to the NN-based predictor and up to $\sim$41\% relative to 802.11ay. This robustness stems from explicit geometric modeling: rather than extrapolating from a beam history, \sysname renders the spectrum at the predicted pose and back-traces the optimal AoD, whereas NN-based predictors break down under rapid changes and reactive scanning fails once the optimal beam shifts beyond its scanning budget. 

\noindent$\blacksquare$ \textbf{Inference Latency.} We evaluate the inference latency for retrieving the optimal Tx beam. The end-to-end neural network ($\sim$5\,ms) is fast but suffers from significant errors. \sysname's full beam-management pipeline for two multiplexed UEs (rendering, back-tracing, and MU-MIMO optimization) completes in $\sim$10.5\,ms per scheduling window (Table~\ref{tab:latency_breakdown}(b)). In contrast, NeRF$^2$ requires over 1\,s, preventing real-time use. (See Appendix~\ref{sec:appendix_latency_analysis}.)

\begin{figure}[t!]
     \centering
     \begin{subfigure}[b]{0.24\textwidth}
         \centering
         \includegraphics[width=\textwidth]{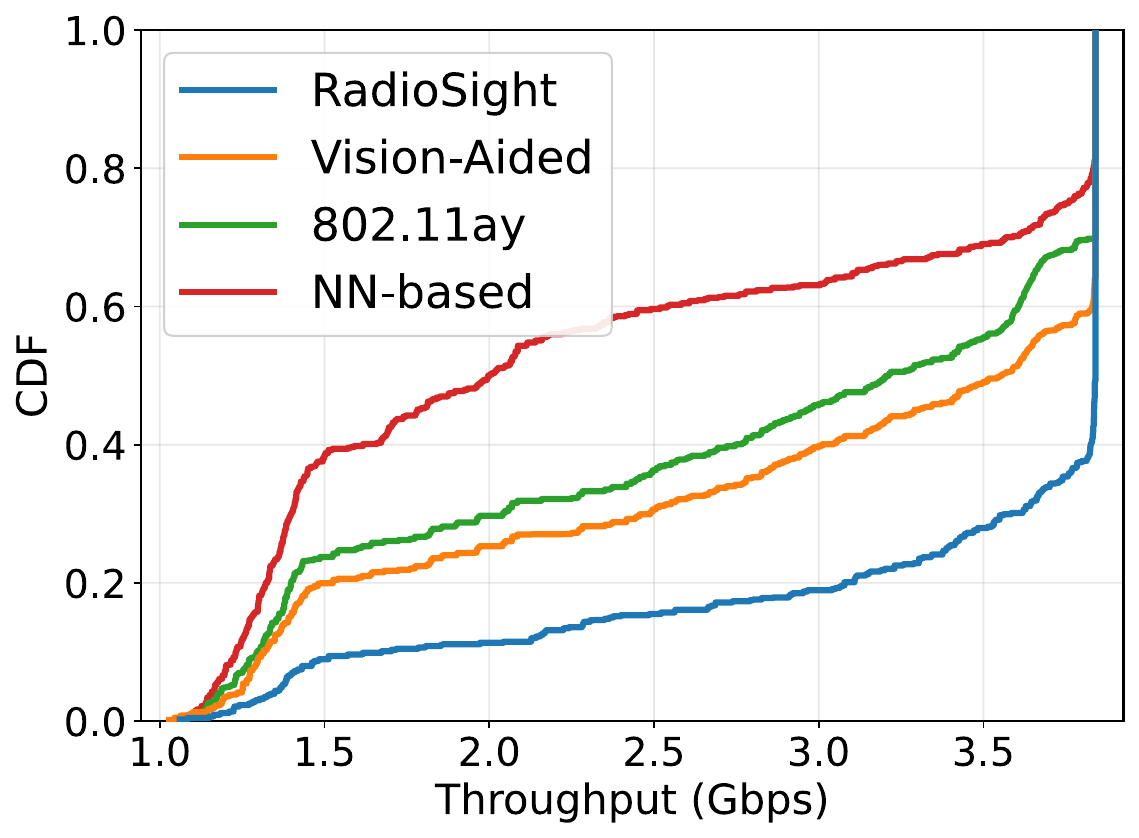}\vspace{-3mm}
         \caption{Throughput CDF.}\vspace{-4mm}
     \end{subfigure}%
     \begin{subfigure}[b]{0.24\textwidth}
         \centering
         \includegraphics[width=\textwidth]{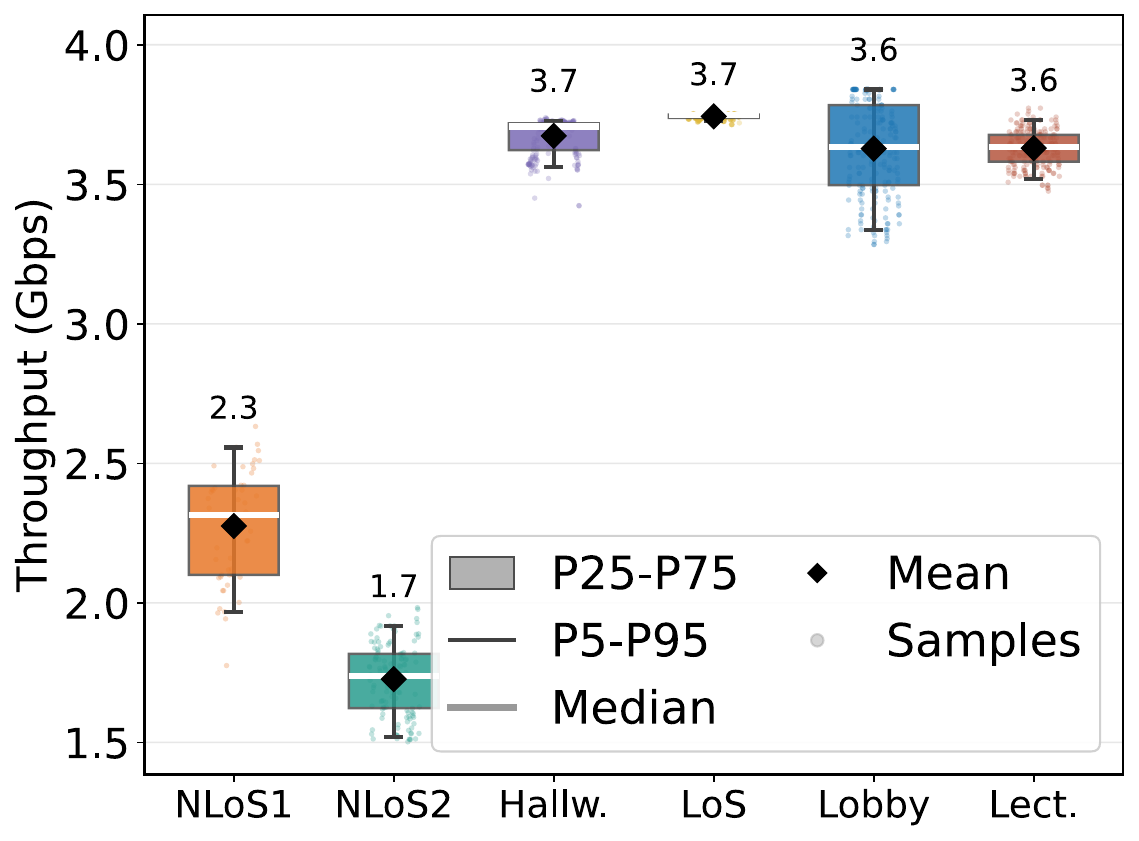}\vspace{-3mm}
         \caption{Per-Scene Throughput.}\vspace{-4mm}
     \end{subfigure}%
     \caption{Throughput Prediction Analysis. \textnormal{(a)~Cumulative distribution of achieved throughput across all evaluated methods. (b)~Achieved throughput of \sysname across diverse scenarios.}}\vspace{-4mm}
     \label{fig:throughput_eval}
\end{figure}

\subsection{End-to-End Throughput Performance}
\label{sec:throughput_prediction}

We next examine how beam-prediction accuracy translates into achieved link throughput.

\noindent $\blacksquare$ \textbf{Throughput Evaluation.} We first evaluate achieved throughput along mobile trajectories, where the user moves between favorable LoS regions and more challenging blocked or NLoS regions. As shown in Fig.~\ref{fig:throughput_eval}(a), \sysname attains the highest median throughput at 3.84\,Gbps versus 3.55\,Gbps for the vision-aided baseline, 3.21\,Gbps for 802.11ay, and 2.00\,Gbps for the NN-based baseline, a $\sim$1.9$\times$ median gain over the NN predictor. The right half of the CDF is relatively compressed because all methods perform similarly when a strong LoS path is available and the link is already close to the PHY ceiling. The separation is largest in the left half of the CDF, which corresponds to blocked or reflective regions where the dominant path changes rapidly: at the 25th percentile \sysname still sustains 3.39\,Gbps, whereas the vision-aided, 802.11ay, and NN baselines fall to 1.97, 1.60, and 1.37\,Gbps, respectively. The vision-aided baseline loses the paths once the UE enters the NLoS region, and the reactive and black-box baselines cannot track the rapidly shifting optimal path; \sysname instead relies on its Radio Field to identify and maintain alternative paths, avoiding the deep throughput drops that pull the baselines' distributions to the left.  These results show that our radio twin is most beneficial precisely in the difficult regions where reactive scanning and black-box prediction fail.

\noindent $\blacksquare$ \textbf{Impact of Scenarios.} We then report per-site throughput across the six evaluated sites. As shown in Fig.~\ref{fig:throughput_eval}(b), \sysname reaches median throughputs of 3.7\,Gbps in both LoS and the hallway, and 3.6\,Gbps in the two large uncontrolled sites, the lobby and the lecture room, while still sustaining 2.3\,Gbps and 1.7\,Gbps in the more challenging NLoS Tx1 and Tx2 settings. Overall, \sysname preserves near-peak throughput across the favorable and large uncontrolled environments and degrades gracefully under blockage, validating the benefit of proactive, geometry-aware network operation across diverse sites.

\begin{figure}[t!]
     \centering
     \begin{subfigure}[b]{0.23\textwidth}
         \centering
         \includegraphics[width=\textwidth]{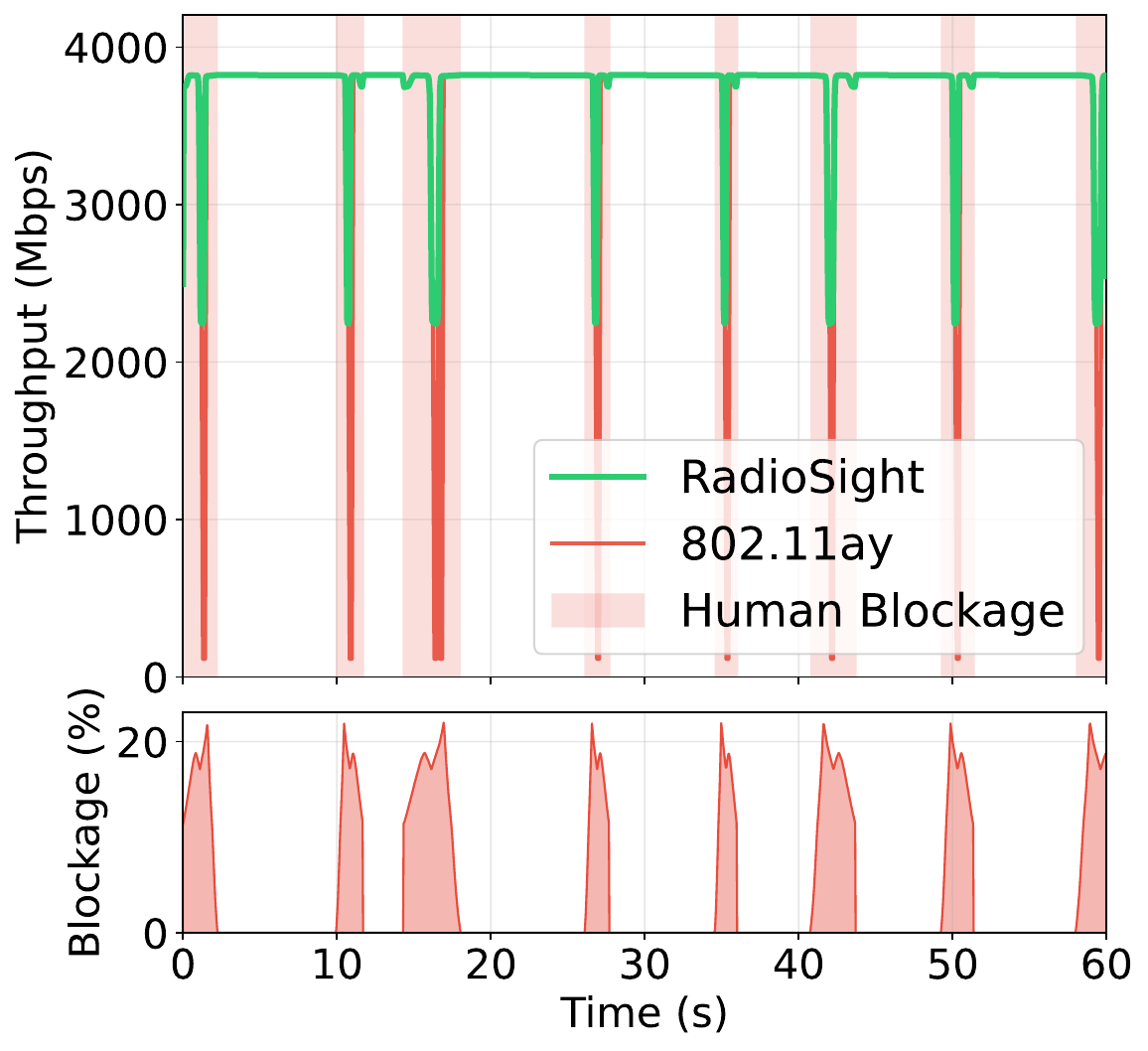}\vspace{-3mm}
         \caption{Dynamic Scene.}\vspace{-4mm}
         \label{fig:dynamic_scene}
     \end{subfigure}%
     \hfill
     \begin{subfigure}[b]{0.23\textwidth}
         \centering
         \includegraphics[width=\textwidth]{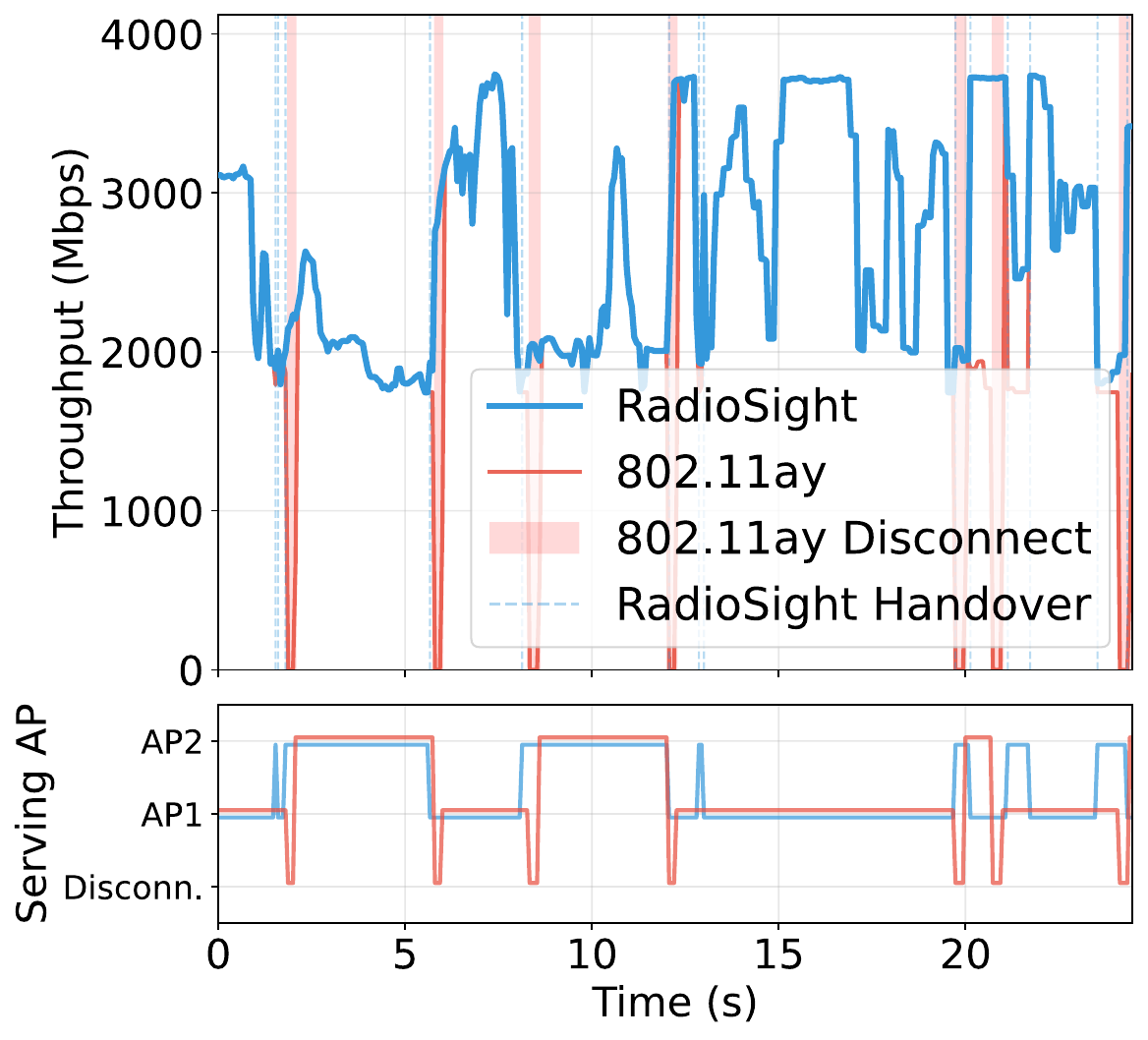}\vspace{-3mm}
         \caption{Multi-AP Handover.}\vspace{-4mm}
         \label{fig:multiuser_qoe}
     \end{subfigure}%
     \hfill
     \caption{Dynamic-Modeling Ablation and Multi-AP. \textnormal{(a)~Ablation of semantic-driven movable object and transient blocker modeling:} \textnormal{performance under dynamic scene changes caused by moving blockers and reflector relocation.} \textnormal{(b)~Performance during multi-AP handovers along mobile trajectories. Unlike reactive baselines that wait for link failure, \sysname maintains continuous connectivity and stable throughput for higher QoE.}}\vspace{-5mm}
     \label{fig:dynamic_performance}
\end{figure}

\begin{figure*}[t!]
     \centering
     \begin{subfigure}[b]{0.24\textwidth}
         \centering
         \includegraphics[width=\textwidth]{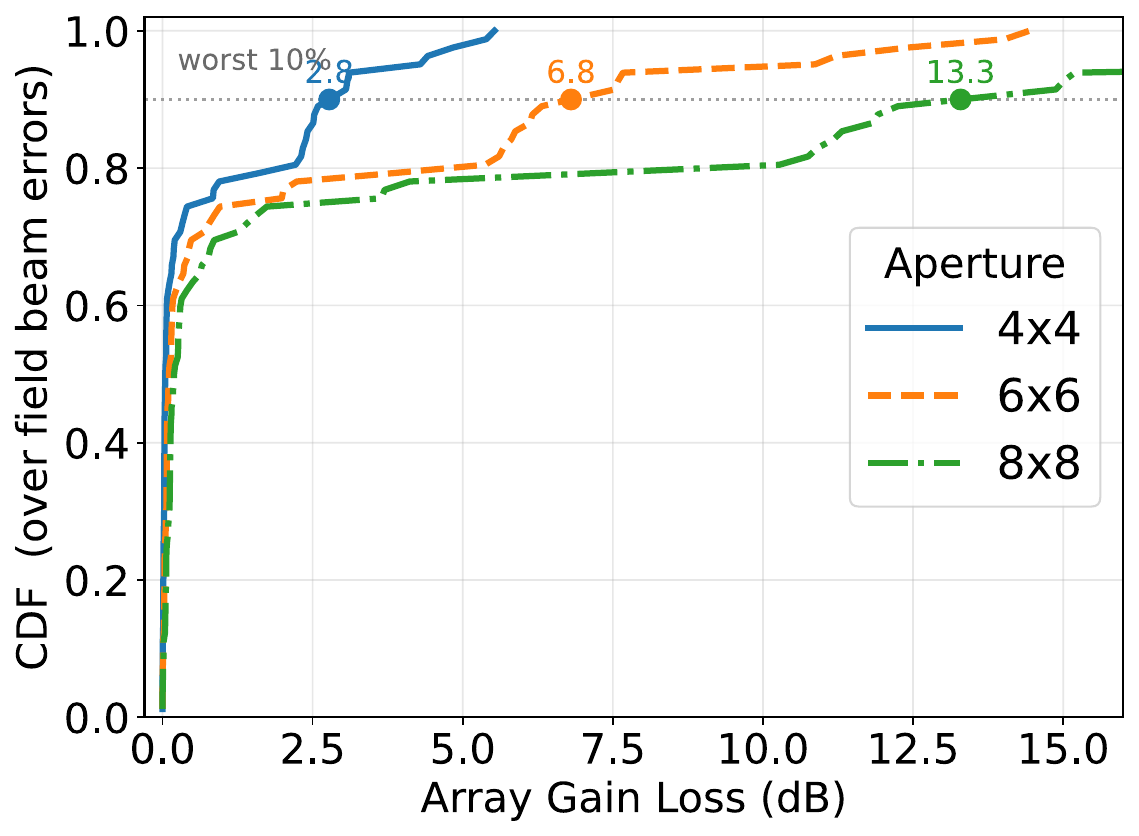}\vspace{-3mm}
         \caption{Cross-Array Generalization.}\vspace{-4mm}
     \end{subfigure}%
     \hfill
     \begin{subfigure}[b]{0.24\textwidth}
         \centering
         \includegraphics[width=\textwidth]{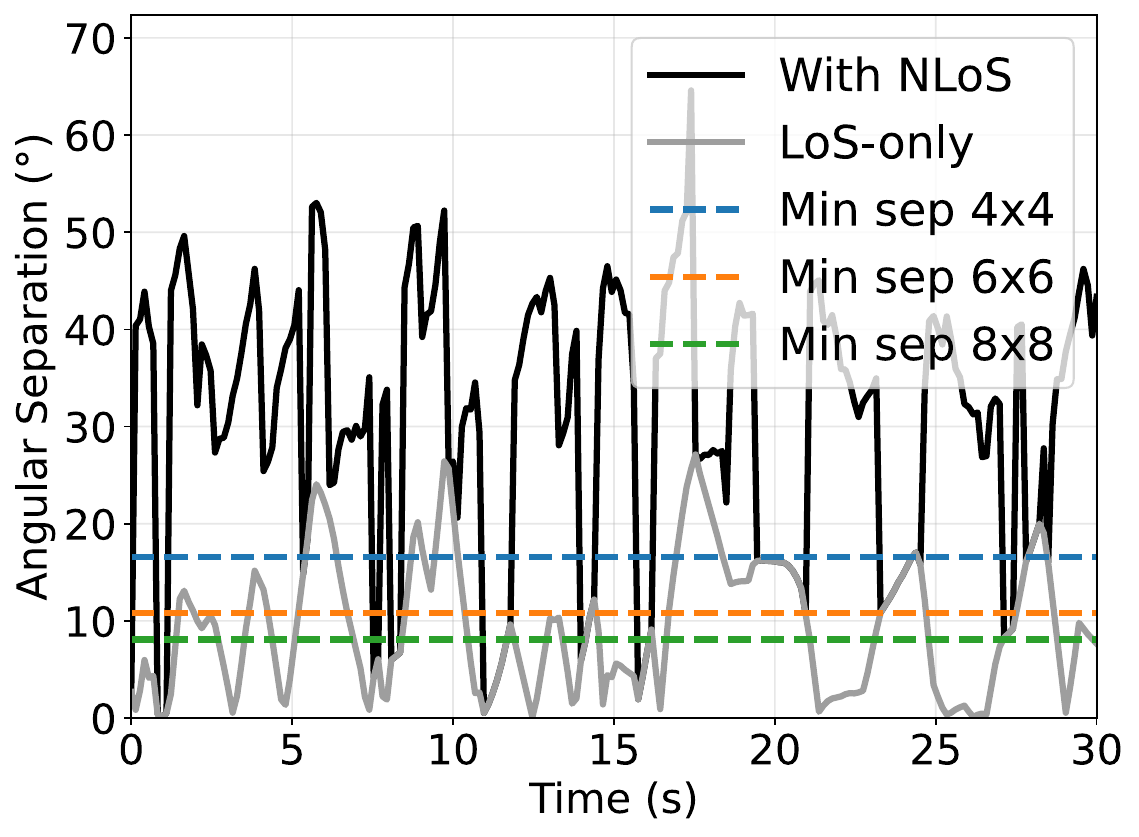}\vspace{-3mm}
         \caption{Beam Separation.}\vspace{-4mm}
     \end{subfigure}%
     \hfill
     \begin{subfigure}[b]{0.24\textwidth}
         \centering
         \includegraphics[width=\textwidth]{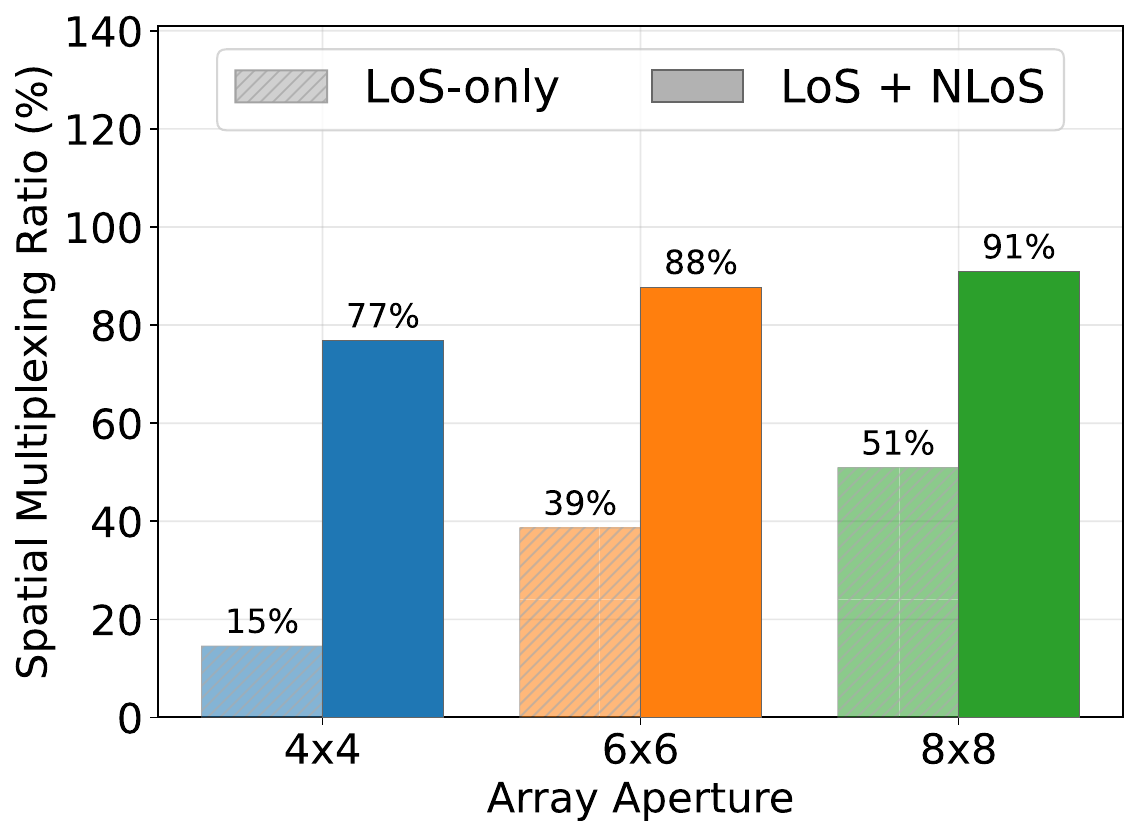}\vspace{-3mm}
         \caption{Spatial Multiplexing Ratio.}\vspace{-4mm}
     \end{subfigure}%
     \hfill
     \begin{subfigure}[b]{0.24\textwidth}
         \centering
         \includegraphics[width=\textwidth]{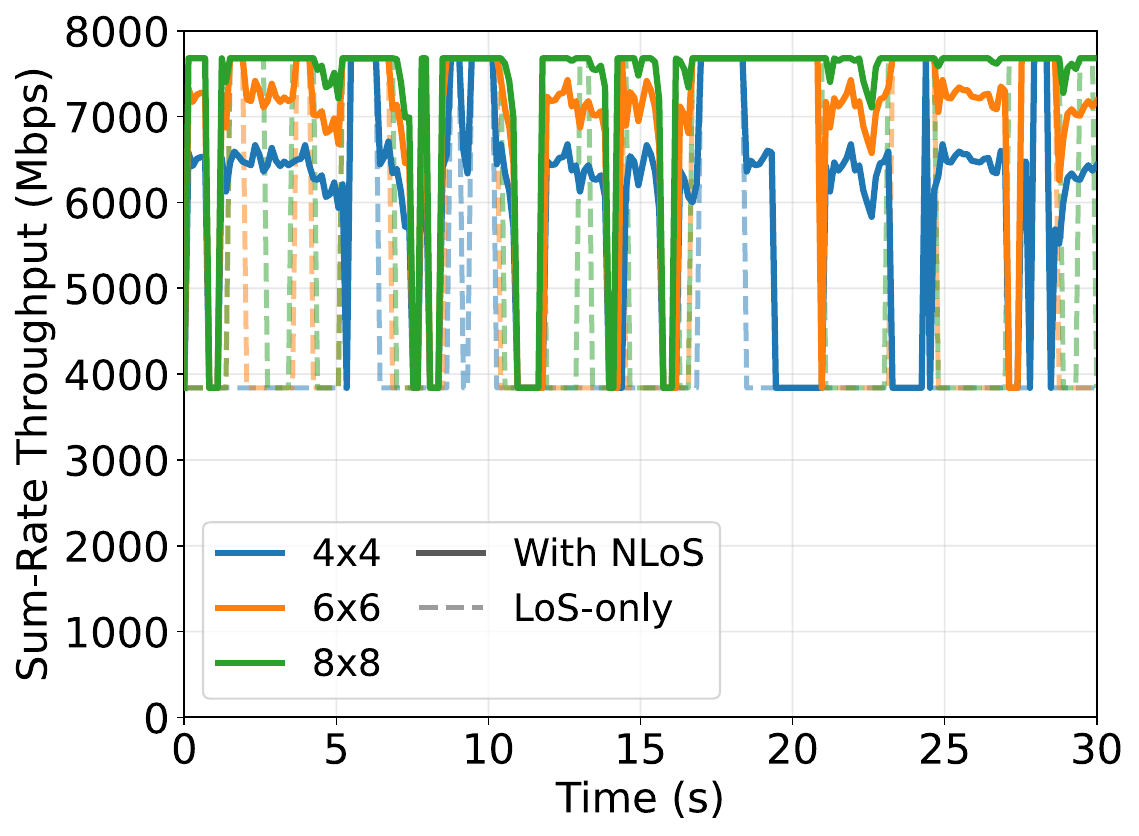}\vspace{-3mm}
         \caption{Achieved Sum-rate.}\vspace{-4mm}
     \end{subfigure}%
     \caption{MU-MIMO and Array Impact. \textnormal{(a)~CDF of the gain loss when the $4\times4$-trained field drives the $4\times4$, $6\times6$, and $8\times8$ apertures. (b)~Dynamic beam separation angle between two UEs. (c)~Proportion of time spatial multiplexing is enabled. (d)~Achieved sum-rate.}}\vspace{-2mm}

     \label{fig:mu_mimo_analysis}
\end{figure*}

\vspace{-2.5mm}\subsection{Dynamics and Handover}\label{sec:dynamics_and_multi_user_scenarios}
\vspace{-0.5mm}
We next evaluate two mobility events that frequently disrupt XR sessions: transient environmental blockage and multi-AP handover (Fig.~\ref{fig:dynamic_performance}). In both cases, \sysname uses the Radio Field to act before the dominant link fails, while scan-based baselines must react only after degradation occurs.

\noindent$\blacksquare$ \textbf{Transient Environmental Blockage.} This experiment serves as our semantic and dynamic-modeling ablation: the reactive baseline represents \sysname without its semantic-driven movable-object modeling (for the relocated metal reflector) and transient-blocker modeling (for the crossing pedestrian). Fig.~\ref{fig:dynamic_scene} shows a pedestrian repeatedly crossing the LoS path while a reflector is relocated. The lower plot reports the blockage ratio over time, and the shaded intervals in the upper plot indicate blockage events. As the LoS path becomes occluded, 802.11ay cannot scan and recover a new path quickly enough, causing throughput to repeatedly collapse to nearly zero. In contrast, \sysname anticipates the blockage and switches to an alternative NLoS path, sustaining much higher throughput in these events. 

\noindent$\blacksquare$ \textbf{Multi-AP Handover.} Fig.~\ref{fig:multiuser_qoe} evaluates a user moving across the coverage regions of multiple APs. The upper plot shows throughput, while the lower plot shows the serving AP over time. \sysname uses the predicted trajectory and Radio Field to pre-schedule AP transitions, producing smooth handovers and avoiding disconnections. By contrast, 802.11ay waits until the current link has already degraded before scanning for a new AP, leading to repeated throughput gaps around AP boundaries.

\vspace{-2mm}\subsection{Impact of Antenna Array}\label{sec:mu_mimo_performance_evaluation}

We examine how array size affects MU-MIMO, and whether a field trained on a small aperture transfers to larger ones. All three arrays are formed from the per-element channels of a single physical $8\times8$ array, and \sysname is trained on the $4\times4$ sub-array's spectrum (Appendix~\ref{sec:appendix_array_spec}).

\noindent\noindent$\blacksquare$ \textbf{Cross-array-size generalization.} Because the field predicts an aperture-independent beam direction, we evaluate its $4\times4$-trained predictions on the larger apertures' measured spectrum (Fig.~\ref{fig:mu_mimo_analysis}(a)). With a median AoD error of only $1.2^\circ$, $78$/$74$/$70\%$ of positions lose under $1$\,dB of array gain on $4\times4$/$6\times6$/$8\times8$; only the worst $10\%$ pays the narrow-lobe penalty ($2.8$/$6.8$/$13.3$\,dB). This indicates that \sysname can drive larger apertures with a small aperture's measurements.

\noindent\noindent$\blacksquare$ \textbf{NLoS modeling expands MU-MIMO.} Two UEs can be spatially multiplexed only when their beams separate beyond the array's $10$\,dB-isolation threshold ($16.6^\circ$/$10.8^\circ$/$8.1^\circ$ for $4\times4/6\times6/8\times8$; Fig.~\ref{fig:mu_mimo_analysis}(b)). By accurately modeling the radio field, \sysname provides each user with alternative NLoS paths, which raise the multiplexing opportunity from $14.5$/$38.6$/$50.9\%$ (LoS only) to $76.8$/$87.7$/$90.9\%$ (LoS+NLoS, Fig.~\ref{fig:mu_mimo_analysis}(c)), lifting the mean sum-rate to $6.0$/$6.9$/$7.3$\,Gbps from $4.4$/$5.3$/$5.8$\,Gbps (Fig.~\ref{fig:mu_mimo_analysis}(d), solid vs.\ dashed). Array size helps through spatial resolution, but it is \sysname's NLoS modeling that turns resolution into reachable multiplexing.

\vspace{-2mm}\section{Case Study: XR Streaming QoE}\label{sec:case_study_mu_mimo_xr_streaming}

To evaluate end-to-end XR streaming, we replay mobile XR sessions in which users walk through the environment while requesting cloud-rendered content. Along each trajectory, we record the delivered throughput, stall events, and motion-to-photon (MTP) latency induced by each method's beam-management decisions. We summarize user-perceived performance using a QoE score, computed as a weighted combination of throughput, stall rate, and MTP latency (See Appendix~\ref{sec:appendix_latency_analysis}). This setup lets us directly test whether better short-horizon prediction translates into smoother XR streaming under mobility.

\begin{figure}[t]
     \centering
          \begin{subfigure}[b]{0.23\textwidth}
         \centering
         \includegraphics[width=0.95\textwidth]{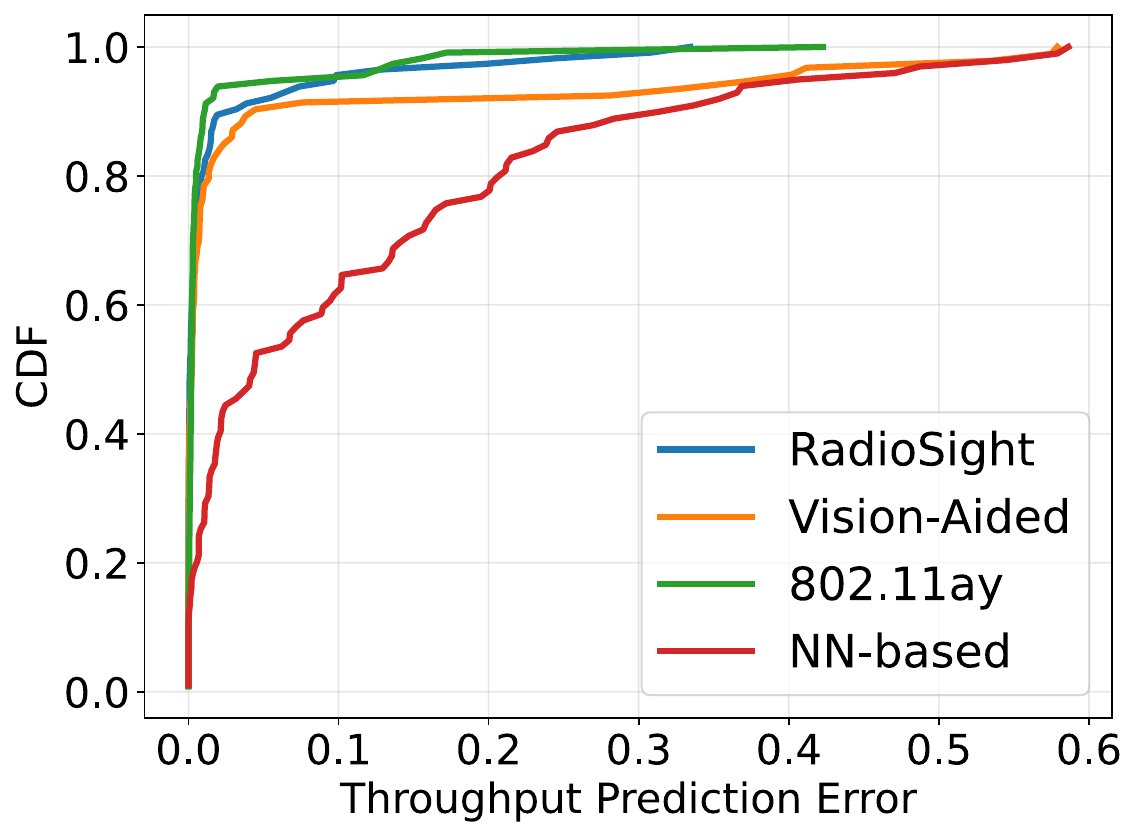}
         \vspace{-3mm}
         \caption{Prediction Error.}\vspace{-4mm}
         \label{fig:qoe_latency} 
     \end{subfigure}%
     \hfill
     \begin{subfigure}[b]{0.23\textwidth}
         \centering
         \includegraphics[width=0.95\textwidth]{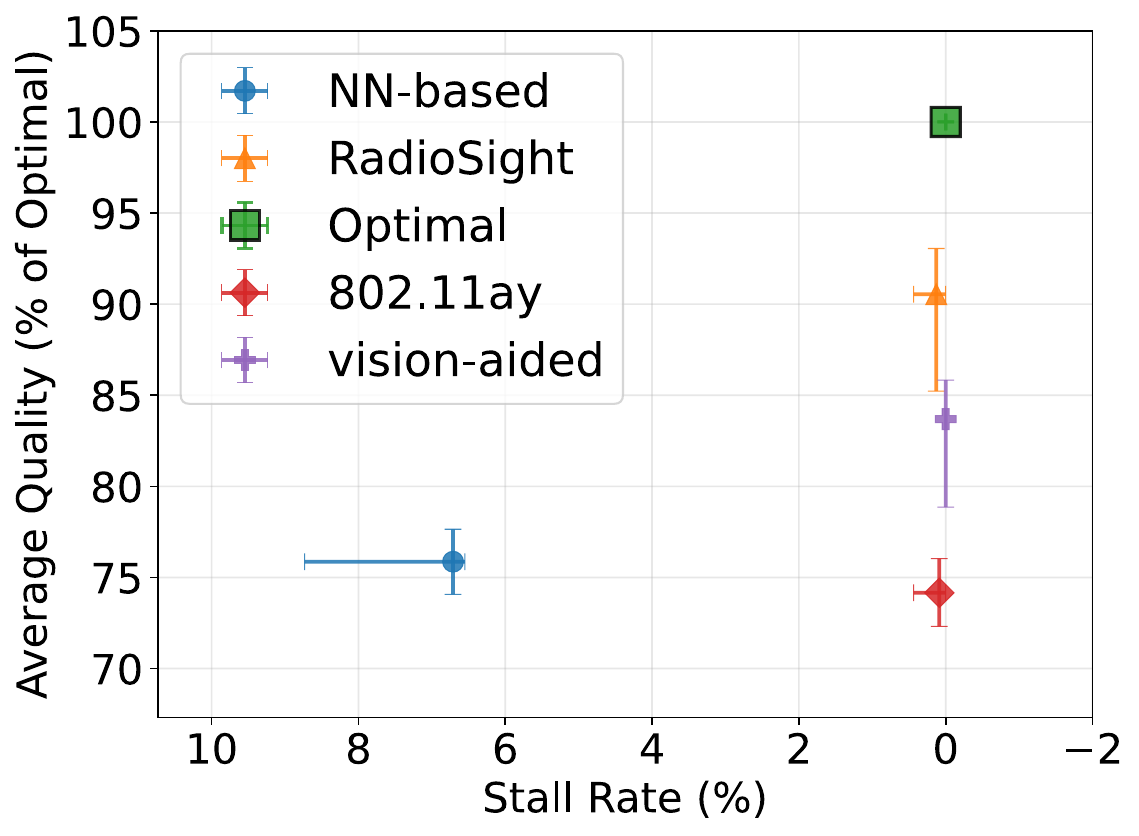}
         \vspace{-3mm}
         \caption{Aggregate QoE.}\vspace{-4mm}
         \label{fig:qoe_general}
     \end{subfigure}
     \caption{Overall XR Streaming QoE. \textnormal{(a)~CDF of throughput prediction errors, which directly affect bitrate selection and streaming stability. (b)~Comparison of aggregate QoE across all methods.}}\vspace{-6mm}
     \label{fig:qoe_evaluation}
\end{figure}

\noindent$\blacksquare$ \textbf{Throughput Prediction.} Accurate throughput prediction is critical for upper-layer streaming control: overestimating future capacity leads to stalls, while underestimating it results in overly conservative transmission and lower visual quality. The throughput-prediction error CDF in Fig.~\ref{fig:qoe_evaluation} shows that \sysname yields the tightest error distribution among the practical baselines. This allows the streaming controller to select more appropriate transmission parameters and maintain a more stable delivery process under mobility.

\noindent$\blacksquare$ \textbf{Achieved XR QoE.} We then evaluate the end-to-end QoE achieved by these different controllers. As shown in Fig.~\ref{fig:qoe_evaluation}, \sysname reaches 82.3\% of the oracle optimum, improving QoE by 55\% over the NN-based method, 25\% over 802.11ay, and 10\% over the vision-aided baseline, while reducing stalls by 3.6$\times$ compared to the NN-based method. These gains stem from more accurate short-horizon throughput prediction together with low-latency beam management, which keeps the XR pipeline responsive during user motion.

\vspace{-2mm}\section{Conclusion}\label{sec:conclusion}
\label{sec:concl}

This paper introduces \sysname, a real-time predictive mmWave optimization system based on multi-modal Radio Fields. By synchronizing the Radio Field with XR-observed dynamics and using backward beam-tracing, \sysname replaces reactive beam scanning with proactive beam optimization under user motion and blockages.

\vspace{-2mm}\begin{acks}
\label{sec:ack}
We appreciate the insightful comments and feedback from the anonymous reviewers and our shepherd. This work was supported in part by the \grantsponsor{NSF}{National Science Foundation}{https://www.nsf.gov/} under Grants No.~\grantnum{NSF}{2521925} and No.~\grantnum{NSF}{2523752}. Zhenlin An is supported by a UGA startup fund.
\end{acks}

\bibliographystyle{ACM-Reference-Format}
\bibliography{references}

\appendix
\section{Details of Multi-modal Radio Field}\label{sec:details_of_multi_modal_radio_field}

\subsection{Segmentation on Multi-modal 3DGS}
\label{segmentation}

In this section, we detail the segmentation process within the trained multi-modal 3DGS model. Once the multi-modal 3DGS backbone is optimized, \sysname performs open-vocabulary 3D scene decomposition to isolate static backgrounds from movable entities. Based on the keywords extracted from the visual input via the VLM, we utilize a frozen CLIP text encoder to extract embeddings for the specific positive vocabulary query (e.g., ``human'' or ``chair'') alongside generic negative vocabularies (e.g., ``objects'' and ``things'').

Specifically, feature splatting appends an additional semantic feature vector $\mathbf{f}_{i} \in \mathbb{R}^{d}$ to each 3D Gaussian. To optimize memory and training efficiency, this feature dimension is heavily compressed to $d=16$. These features are rendered in a view-independent manner, as the intrinsic semantics of an object remain consistent regardless of the viewing direction. To bridge the dimensional gap between the compact rendered features and the high-dimensional CLIP space, a shallow MLP processes the rendered intermediate features to output the final CLIP-aligned feature maps.

For each Gaussian in the scene, we compute the pairwise cosine similarity between its projected semantic feature and the encoded text embeddings. By applying a temperature-scaled softmax to these similarity scores, we generate a probability distribution over the vocabularies. Gaussians whose similarity to the positive target query exceeds a predefined confidence threshold (e.g., $\tau = 0.6$) are explicitly segmented from the background. This allows \sysname to register and track them as distinct, independent physical obstacles in the digital twin.

\subsection{Multi-modal Training}
\label{sec::gaussian_regul}

\textbf{Geometric Regularization: }To further motivate the geometric regularizations introduced in Section~\ref{ss:radio_scene_registration}, it is critical to understand the fundamental limitations of relying exclusively on LiDAR point cloud initialization combined with radio-spectrum-only training. While LiDAR sensors provide accurate absolute depth measurements, the resulting point clouds are inherently sparse and discrete. Unless the environment is subjected to extremely dense scanning or computationally heavy post-processing (e.g., Poisson surface reconstruction), raw LiDAR data lacks explicit, continuous surface information. Furthermore, radio spectrum is intrinsically noisy and heavily distorted by conventional beamforming (CBF) lobe widths and side lobe interference. 

Consequently, initializing a 3DGS representation with sparse LiDAR and optimizing it strictly using noisy radio measurements yields a highly ill-posed optimization problem. Without explicit geometric constraints, the unconstrained Gaussians will attempt to fit the ambiguous RF interference patterns by expanding into volumetric, ``fluffy'' light clusters rather than converging to rigid physical boundaries. This geometric degradation is catastrophic for deterministic radio propagation modeling. Accurate physical ray back-tracing strictly relies on well-defined solid surfaces to compute valid angles of incidence, reflections, and diffractions. If the scene geometry is represented by ambiguous, cloud-like Gaussians, the estimated radio-surface intersections become entirely unreliable, breaking the underlying ray-tracing physics.

To suppress this volumetric ambiguity and enforce the formation of continuous, rigid surfaces, we introduce two explicit geometric regularization terms: a scale penalty $\mathcal{L}_{scale}$ and a normal alignment loss $\mathcal{L}_{normal}$, which collectively form the regularization objective $\mathcal{L}_{reg} = \lambda_s \mathcal{L}_{scale} + \lambda_n \mathcal{L}_{normal}$.

First, to ensure the Gaussians physically deform into flat, surfel-like disks that cleanly represent physical boundaries rather than intersecting volumes, we apply a scale regularization. This loss directly penalizes the magnitude of the shortest axis of each Gaussian, encouraging it to approach zero:
\begin{equation}\small
\mathcal{L}_{scale} = \sum_{i=1}^{N} \min(s_{i,1}, s_{i,2}, s_{i,3})
\end{equation}
where $s_{i,1}, s_{i,2}, s_{i,3}$ represent the diagonal elements of the scaling matrix $S_i$ for the $i$-th Gaussian. Once the Gaussians are flattened into surfels, their shortest axis naturally defines the local microscopic surface normal. We define this per-Gaussian normal $\mathbf{n}_i$ as the column vector of the rotation matrix $R_i$ corresponding to the minimized scale value $\min(s_{i,1}, s_{i,2}, s_{i,3})$. 

With the microscopic normals defined, we enforce consistency between these individual surfels and the macroscopic scene geometry. We extract the macroscopic surface normal $\mathbf{N}_{D}$ directly from the spatial gradients of the rendered depth map $\hat{\mathbf{D}}$. Concurrently, the predicted normal map $\hat{\mathbf{N}}$ is volume-rendered using the standard $\alpha$-blending formulation:
\begin{equation}\small
\hat{\mathbf{N}} = \sum_{i=1}^{N} \mathbf{n}_i \alpha_i T_i
\end{equation}
The normal alignment loss forces the rendered per-Gaussian normals to closely align with the macroscopic surface geometry derived from the depth map $\hat D$:
\begin{equation}\small
\mathcal{L}_{normal} = 1 - \mathbf{N}_{\hat D} \cdot \hat{\mathbf{N}}
\end{equation}

The effectiveness of this geometry regularization is strongly validated by our empirical results. As demonstrated in the evaluation section, the baseline method WRF-GS, which relies on radio-spectrum-only training with LiDAR initialization, exhibits severely degraded performance. In contrast, by enforcing the $\mathcal{L}_{scale}$ and $\mathcal{L}_{normal}$ constraints to flatten the Gaussians into distinct surfels and align their orientations, \sysname successfully suppresses volumetric ambiguity, ensures geometrically accurate physical boundaries, and significantly outperforms WRF-GS in high-fidelity radio field reconstruction.

\subsection{Single-Bounce Energy Fraction}
\label{sec:single_bounce_energy} 
To support the single-bounce assumption used in \S\ref{ss:beam_prediction}, we validate it across a diverse set of indoor environments using both real-world measurements and ray-tracing simulations. We quantify the fraction of resolvable NLoS energy contributed by multi-bounce paths, reported as the mean and 90th percentile (P90) across Rx positions.
Our measurements are collected in two uncontrolled real-world environments: a large, relatively open $16.4 \times 16.2$\,m lobby and a furnished $21.0 \times 10.2$\,m lecture room. We further simulate seven representative indoor environments spanning different room types, scales, and scene complexities: a digital twin of the measured lobby and six furnished rooms from the WiSegRT dataset~\cite{zhang2024wisegrt}. All simulations are performed in Sionna RT at 28\,GHz with up to four interactions per path. The six WiSegRT environments include a living room of $6.2 \times 4.2 \times 2.4$\,m with 33 objects, a bedroom of the same size with 24 objects, a studio of $6.0 \times 4.0 \times 2.4$\,m with 28 objects, a classroom of $6.2 \times 4.2 \times 2.4$\,m with 46 objects, an open-plan office of $16.0 \times 8.0 \times 2.4$\,m with 159 objects, and a storage hall of $9.5 \times 5.0 \times 2.4$\,m with 43 objects whose surfaces are 64\% metal. For each environment, we resolve the channel at each Rx position into distinct geometrical paths.

Table~\ref{tab:bounce_energy} summarizes the results. In open scenes, single-bounce paths account for almost all of the NLoS energy. Multi-bounce paths contribute 1.8\% of the NLoS energy in the measured lobby and 2.4\% in its digital twin, which confirms that the simulation reproduces the measurement in the same room. This contribution rises to 7.3\% in the furnished lecture room and to between 9.8\% and 16.4\% in the six furnished rooms. The P90 values show that the effect concentrates at a minority of positions, where multi-bounce paths carry 25.5\% to 46.4\% of the NLoS energy. These results indicate that the single-bounce-only assumption is well justified for open indoor environments and remains a reasonable approximation for furnished rooms. Nevertheless, incorporating multi-bounce propagation in future work would improve modeling fidelity in small and densely furnished rooms, and in particular at the minority of positions where higher-order interactions dominate.

\begin{table}[t]
  \centering
  \caption{Multi-bounce share of the resolvable NLoS energy, per RX position (mean and P90 across positions).}\vspace{-2mm}
  \label{tab:bounce_energy}
  \small
  \setlength{\tabcolsep}{5pt}
  \begin{tabular}{@{}lccc@{}}
    \toprule
    \textbf{Scene} & \textbf{Source} & \textbf{Mean (\%)} & \textbf{P90 (\%)} \\
    \midrule
    Lobby (open)              & measured  & 1.8  & 4.7  \\
    Lecture room (furnished)  & measured  & 7.3  & 19.2 \\
    \midrule
    Lobby (open)              & simulated & 2.4  & 5.7  \\
    Living room               & simulated & 9.8  & 25.5 \\
    Bedroom                   & simulated & 12.2 & 33.8 \\
    Studio                    & simulated & 14.3 & 38.4 \\
    Classroom                 & simulated & 16.4 & 38.2 \\
    Open-plan office          & simulated & 15.6 & 46.4 \\
    Storage hall              & simulated & 11.4 & 37.0 \\
    \bottomrule
  \end{tabular}\vspace{-4mm}
\end{table}

\subsection{Object Dynamic Modeling Details}\label{sec:short_term_object_dynamic_modeling}
This section provides additional technical details regarding the implementation of the dynamic twin.

\textbf{Client Local Perception and Dynamic Detection.} 
The client leverages a high-frequency IMU (e.g., 1000\,Hz) and visual-inertial odometry (VIO) to provide robust 6DoF pose estimation. By employing a kinematic motion model, the UE performs pose prediction over a 100\,ms horizon, maintaining roughly 1--2\,cm position error and a few degrees of orientation error (Fig.~\ref{fig:pose_ekf_horizon})---a precision level sufficient to compensate for the end-to-end wireless feedback loop. 

\textbf{Local Geometry and Server Synchronization.} The local geometry is maintained by the client's own SLAM service and drives detection on its own; the server is an information source for it, not its owner. The server holds the same scene as a lightweight 3D bounding-box abstraction alongside the multi-modal field, assigns each movable box an EKF state, and pushes updates over the control channel so that every client converges to a consistent view. It builds those updates by associating the 2.5D proposals arriving from all clients with the boxes it already tracks and fusing each match into that box's EKF state~\cite{weng2020ab3dmot}. 

\textbf{UE-Side Depth Discrepancy Detection.} As shown in Fig.~\ref{fig:object_registration}, the headset relies on its local perception to detect environmental changes without constantly querying the server. The client headset maintains a lightweight, local geometry proxy representing the known static environment. Utilizing its onboard 6DoF tracking, the UE generates a continuous sequence of ``Local Geometry Depth Maps'' (denoted as $D_{proxy}$). Simultaneously, the headset's active sensors (e.g., structured light or stereo cameras) capture the ``Online Depth Map Flow'' ($D_{real}$) representing the actual, real-time physical space. 

By performing a pixel-wise differencing between the expected local depth and the actual online depth, the UE can immediately isolate newly introduced or moved objects in its field of view through a pixel-level discrepancy check:
\begin{equation}
    \Delta D = |D_{real} - D_{proxy}| > \delta_{th}
\end{equation}
To robustly estimate the object's extent from noisy structured-light data and isolate the true target, we employ a \textit{Depth-Probability Weighted Centroid} method. For clusters exceeding the threshold $\delta_{th}$, we generate a 2.5D proposal. To minimize the uplink overhead, the client does not stream the full point cloud; instead, it extracts a simple 2D bounding box (shown as the red rectangle in Fig.~\ref{fig:object_registration}) alongside a characteristic depth $d_c$ derived from the mode of the depth histogram, and transmits this sparse geometric proposal to the edge server.

\textbf{Server Global Dynamic Fusion and Registration.} 
The server consolidates these sparse 2.5D proposals into a globally consistent 3D state. Upon receiving a proposal packet (timestamp, 6DoF pose, 2.5D box), the server back-projects the coordinates to form a 3D candidate $\mathcal{B}_{cand}$. Following the \textit{Tracking-by-Detection} paradigm of AB3DMOT~\cite{weng2020ab3dmot}, we implement a \textit{Bayesian Centroid Fusion} pipeline to categorize the dynamics:

\begin{itemize}[leftmargin=*]
    \item \textit{Object Movement:} If the candidate $\mathcal{B}_{cand}$ is associated with an existing entity via 3D IoU and Mahalanobis distance, the server updates its translation and rotation. An Extended Kalman Filter (EKF) maintains the state vector $\mathbf{x} = [x, y, z, \theta, w, l, h, \dot{x}, \dot{y}, \dot{z}]^T$. The position of each connected UE is also aligned with a kinematic human model, incorporating its predicted future trajectory.
    \item \textit{New Entity Discovery:} If no association is found, the server initializes a tentative track. We adopt an $M$-out-of-$N$ consistency check (e.g., 5 out of 8 frames) to confirm the existence of new objects and filter out transient noise. Once confirmed, the entity is registered as a \textit{Short-term Obstacle}, parameterized by a 3D bounding box with the aforementioned state vector to model its mobility. It is assigned a high RF opacity to provide immediate, conservative blockage modeling for radio paths.
    \item \textit{Entity Solidification:} If a short-term obstacle remains stationary for a threshold duration, it is promoted to a \textit{Registered Object}, triggering localized Gaussian training to distill its specific visual and RF features.
    \item \textit{Entity Deletion:} Removes entities that have exited the field of view or become mismatched from their original tracks. When a registered object moves too far to maintain its tracking association, the system treats it as a new short-term obstacle and solidifies it in the new position later, subsequently discarding the stale original entity to maintain scene consistency.    \item \textit{Updating UE Local Geometry:} The server periodically dispatches the simplified, fused global geometry back to the UEs, providing them with an updated global context for future depth discrepancy checks.
    \item \textit{Multi-modal Feedback:} To continuously update the movement and radio appearance of existing long-term objects, the UE monitors its relative pose. Once the pose change exceeds a predefined threshold, the UE triggers a supplementary update to transmit an aligned visual frame coupled with a low-resolution radio spectrum measurement.
\end{itemize}

This tiered architecture allows \sysname to track dynamic blockers with $\sim$2.3\,ms processing latency on the client, while maintaining a high-fidelity radio field on the server that dynamically evolves with the physical environment.

\subsection{MU-MIMO Beamforming}
\label{sec:mu-mimo_beamforming}
This subsection provides additional analysis and technical details regarding the MU-MIMO.
\textbf{Optimization problem formulation:}
As described in the main text, the query-and-trace pipeline returns a set of fully resolved multipath candidates for each UE $u$ at time $t$:
\begin{equation}\small
\mathcal{C}_{u,t} = \{[\text{AoD, AoA, RSSI}]\}_{1 \dots K} = \text{QueryRF}_{\mathcal{M}_{GS}}(\mathbf{P}_{Rx,u}, \Phi_{Rx,u})
\end{equation}
where $\mathbf{P}_{Rx,u}$ denotes the predicted pose and $\Phi_{Rx,u}$ is derived from the hardware's configuration.

Leveraging these deterministic multipath profiles, the edge server formulates a joint MU-MIMO user scheduling, frequency allocation, and beamforming problem. Let $\mathcal{F}$ denote the set of available orthogonal frequency sub-channels (e.g., $|\mathcal{F}| = 16$). We introduce a binary variable $x_{u,k,f} \in \{0, 1\}$ to indicate whether the $k$-th candidate beam from $\mathcal{C}_{u,t}$ is selected as the primary communication path for UE $u$ on sub-channel $f$. Consequently, $y_{u,f} = \sum_{k=1}^K x_{u,k,f}$ acts as an indicator of whether UE $u$ is admitted into the spatial multiplexing group operating on sub-channel $f$. 

To isolate highly correlated paths and maximize the network sum-rate while adhering to strict XR QoE requirements, we formulate this as a Mixed-Integer Non-Linear Programming (MINLP) problem~\cite{klanvsek2015comparison}:

\begin{subequations}\small
\begin{align}
\max_{\mathbf{X}, \mathbf{W}} \quad & \sum_{f \in \mathcal{F}} \sum_{u \in \mathcal{U}} y_{u,f} \log_2(1 + \text{SINR}_{u,f}) \label{eq:obj} \\
\text{s.t.} \quad
& \sum_{f \in \mathcal{F}} y_{u,f} \log_2(1 + \text{SINR}_{u,f}) \ge \left( \sum_{f \in \mathcal{F}} y_{u,f} \right) R_{\text{req}}, \quad \forall u \in \mathcal{U}, \label{eq:qos} \\
& \sum_{u \in \mathcal{U}} y_{u,f} \le N_{\text{RF}}, \quad \forall f \in \mathcal{F}, \label{eq:rf_chains} \\
& \sum_{f \in \mathcal{F}} \sum_{u \in \mathcal{U}} y_{u,f} \|\mathbf{w}_{u,f}\|^2 \le P_{\max}, \label{eq:power} \\
& \sum_{f \in \mathcal{F}} \sum_{k=1}^K x_{u,k,f} \le 1, \quad \forall u \in \mathcal{U}, \label{eq:single_assignment} \\
& y_{u,f} = \sum_{k=1}^K x_{u,k,f}, \quad x_{u,k,f} \in \{0, 1\}, \label{eq:binary}
\end{align}
\end{subequations}
where $R_{\text{req}}$ is the minimum required data rate, $N_{\text{RF}}$ is the maximum number of supported spatial streams per sub-channel (dictated by the RF chain limit), and $P_{\max}$ is the total available transmit power budget across the entire array and spectrum. 

Constraint \eqref{eq:single_assignment} ensures that a UE is assigned to at most one specific candidate beam and one specific frequency group to maintain hardware simplicity. Under this orthogonal frequency division, inter-user interference only occurs between UEs co-scheduled on the \textit{same} sub-channel $f$. Thus, the SINR for a scheduled UE $u$ on sub-channel $f$ is given by:
\begin{equation}
\text{SINR}_{u,f} = \frac{|\mathbf{h}_{u,f}^H \mathbf{w}_{u,f}|^2}{\sum_{v \in \mathcal{U} \setminus \{u\}} y_{v,f} |\mathbf{h}_{u,f}^H \mathbf{w}_{v,f}|^2 + \sigma^2}
\end{equation}

\textbf{Suboptimal solution for practice:}
Solving the above MINLP globally is NP-hard and computationally intractable for the latency budget of XR applications, particularly in dense scenarios with numerous UEs. Therefore, we propose a tiered, sub-optimal heuristic algorithm that decouples frequency-domain user grouping from spatial beam synthesis:

\begin{enumerate}[leftmargin=*]
    \item \textbf{Inter-Group Orthogonalization (Frequency Division):} In high-density scenarios, the available array resources and spectrum are partitioned into multiple orthogonal frequency division multiple access (FDMA) groups. To proactively prevent severe spatial interference, UEs exhibiting highly correlated or overlapping angle of departure (AoD) paths are explicitly separated and assigned to different frequency groups. This step ensures baseline isolation before spatial multiplexing is attempted.
    \item \textbf{Intra-Group Greedy Scheduling (Spatial Sectorization):} Within each individual FDMA group, the algorithm falls back to a greedy strategy to finalize sub-optimal spatial multiplexing assignments. UEs are sorted based on the RSSI of their top candidate beams. To minimize computational complexity, the transmitter's field of view (FoV) is discretized into uniform spatial sectors bounded by the sub-main-lobe width. The AoD of each candidate beam is mapped directly to a sector index. The algorithm iteratively admits UEs into the group's spatial schedule, provided their selected beams map to unoccupied sectors. This $O(1)$ index lookup rapidly guarantees spatial separation without relying on complex floating-point angular distance evaluations.
    \item \textbf{Constraint Assignment and Fast LCMV Synthesis:} For the scheduled UEs within a group, the selected candidate's AoD dictates the main lobe direction $\theta_{u}^{\text{main}}$. To suppress intra-group cross-talk, the AoDs of all \textit{other} co-scheduled UEs in the same FDMA group ($v \neq u$) are assigned as explicit discrete null-steering constraints: $\mathcal{S}_{\text{null}, u} = \{ \text{AoD}_{v,j} \mid x_{v,j} = 1, v \neq u \}$. The final weight vector $\mathbf{w}_u$ is then rapidly synthesized using the closed-form Linearly Constrained Minimum Variance (LCMV) solution~\cite{buckley1987spatial}:
    \begin{equation}
    \mathbf{w}_u = \mathbf{C}_u(\mathbf{C}_u^H \mathbf{C}_u + \epsilon \mathbf{I})^{-1} \mathbf{f}
    \end{equation}
    where $\mathbf{C}_u$ contains the steering vectors for $\theta_{u}^{\text{main}}$ and $\mathcal{S}_{\text{null}, u}$, and $\epsilon$ is a regularization parameter. Because the spatial multiplexing size $N_{\text{RF}}$ per group is strictly bounded, the constraint matrix dimension remains extremely small. The resulting matrix inversion is practically instantaneous: the entire MU-MIMO beam-synthesis pipeline completes in $0.24$\,ms for $K=2$ co-scheduled UEs (Table~\ref{tab:latency_breakdown}(b)).
\end{enumerate}

\begin{figure*}[t!]
    \centering
    \begin{subfigure}[b]{0.48\textwidth}
        \centering
        \includegraphics[width=0.95\textwidth]{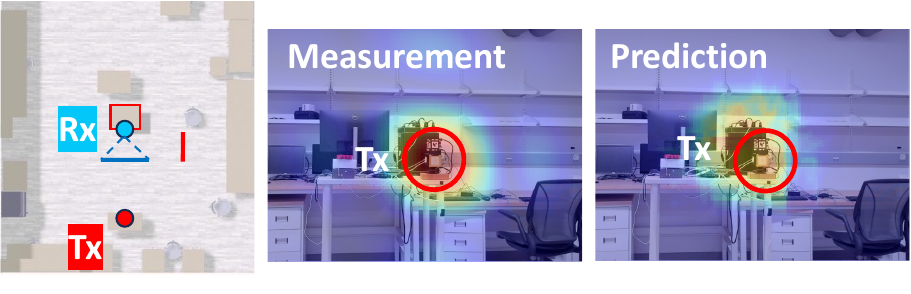}\vspace{-4mm}
        \caption{LoS Measurement vs. Prediction.}\vspace{-3mm}
    \end{subfigure}
    \hfill
    \begin{subfigure}[b]{0.48\textwidth}
        \centering
        \includegraphics[width=0.95\textwidth]{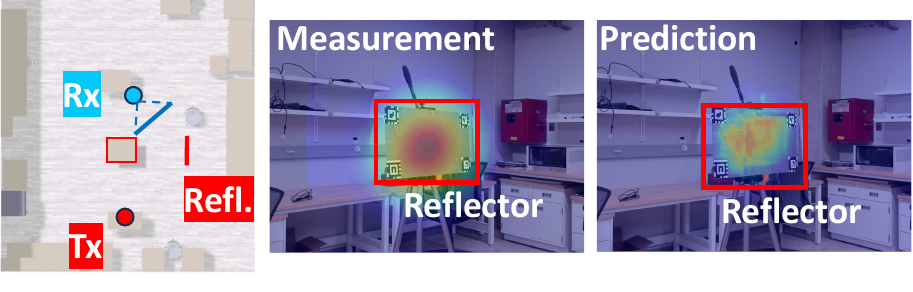}\vspace{-4mm}
        \caption{NLoS Measurement and Prediction.}\vspace{-3mm}
    \end{subfigure}
    \caption{Radio Field based Beam Prediction. \textnormal{(a)~LoS scenario: the floor plan (left) shows direct Tx--Rx alignment; the measured (middle) and predicted (right) spatial spectra both concentrate energy around the Tx direction. (b)~NLoS scenario: a metal reflector redirects the signal; the measured spectrum (middle) shows a dominant peak at the reflector, which the Radio Field prediction (right) faithfully reproduces. The close agreement validates the fidelity of \sysname's spatial spectrum synthesis under both propagation regimes.}}\vspace{-5mm}
    \label{fig:spectrum_similarity}
\end{figure*}

\section{Evaluation}\label{sec:radio_field_evaluation}
This Appendix provides complementary results and discussions for \S\ref{sec:evaluation}.

\subsection{Beam Scanning Codebook Construction}
\label{app:codebook}
A static beam scanning codebook was designed to balance three competing factors: angular steering resolution, the beam width imposed by the $4 \times 4$ testbed radio-head aperture, and the time required to complete the densest possible scan at each scene pose.
During data collection, a Python generator function was iterated until exhausted, ensuring each steering command synchronized with SDR packet transmission. Two nested generator loops (one for Tx and one for Rx) produced a spatial-raster codebook consisting of 17 Tx and 55 Rx beam combinations. Each codebook covers a $90^\circ$ FoV with $45^\circ$ elevation range and $360^\circ$ azimuth range. The Tx codebook included an elevation spacing of $22.5^\circ$, and an azimuth spacing of $45^\circ$. The Rx codebook was denser with an elevation spacing of $15^\circ$, and an azimuth spacing of $20^\circ$.
Each radio chain utilized a local GNU Radio process and USB 3.0 to execute packet encoding and decoding via an 802.11 out-of-tree (OOT) library. Control channel access for GNU Radio utilized ZeroMQ (ZMQ) sockets. All scanning commands were initiated from the Rx PC over the control channel. Within the beam steering generator loops, the Rx PC commanded the Tx SDR via ZMQ to transmit a packet. The Rx PC then handled packet decoding, temporary storage, and timing alignment of the spatial-CSI data synchronized with the Aria glasses.

\subsection{Training and Rendering Details}
Each Gaussian's directional radio radiance is modeled with spherical harmonics of degree $3$, activated one order every $1000$ steps. For evaluation, fields are optimized with Adam for up to $30$k iterations and train within the $24$\,GB of a single RTX~4090; for cold start we adopt a shortened schedule ($7$k iterations with adjusted training setup) that completes in $\sim$1--2\,min. At inference the field is queried on the receiver pixel grid $\mathcal{C}_{rx}$, with the rendering resolution reduced to match the array beamwidth.

\subsection{Spectrum Prediction Similarity}
\label{sec:spectrum_prediction_similarity}

We omit the detailed evaluation of spectrum prediction similarity from the main text, as it is orthogonal to our primary end-to-end performance metrics. However, evaluating the fidelity of the rendered spectrum provides valuable insight into the accuracy of our spatial modeling. Specifically, our rendered spectrum achieves a PSNR of 16.7, an SSIM of 0.302, and an LPIPS of 0.287 when compared against the ground truth. Visual examples of these rendering results are illustrated in Fig.~\ref{fig:spectrum_similarity}. While the absolute values of these metrics are notably lower than those typically observed in standard computer vision (CV) tasks, this discrepancy is expected. The ground truth radio spectrum is inherently noisy and heavily affected by interference, primarily due to the strong side lobes generated by CBF on our hardware platform. Despite these lower absolute scores, the results demonstrate strong modeling capability; the rendered spectrum accurately captures and represents the spatial locations of the dominant radio paths within the 3D environment.

\begin{table}[t]
  \centering
  \caption{Per-stage latency of \sysname's three concurrent threads, at a unified 120\,Hz. \textnormal{\emph{M}~=~measured on this RTX~4090 server, \emph{A}~=~analytic, \emph{B}~=~a standard 5G-NR / render budget. Panels (a,\,c) give the deployment budget (air-interface network hops); panel (d) the motion-to-photon chain on our current prototype test. The threads run concurrently, so their totals are independent and not additive; per-stage rationale in \S\ref{sec:appendix_latency_analysis}.}}\vspace{-2mm}
  \label{tab:latency_breakdown}
  \footnotesize
  \setlength{\tabcolsep}{4pt}
  \begin{tabular}{@{}lrc@{}}
    \toprule
    \textbf{Stage} & \textbf{Latency (ms)} & \textbf{Type} \\
    \midrule
    \multicolumn{3}{@{}l}{\textbf{(a) Scene-update thread} ($T_{\text{update}}$, deployment)} \\
    \midrule
    Blocker detection (UE)              & 2.174 & M \\
    Update-packet serialization         & 0.009 & M \\
    Uplink over the air (5G-NR CG)      & 1.500 & B \\
    Packet deserialization (server)     & 0.007 & M \\
    Multi-view fusion (server)          & 0.015 & M \\
    Blocker EKF update (server)         & 0.052 & M \\
    \cmidrule(r){1-3}
    \textbf{Total} $T_{\text{update}}$  & \textbf{3.757} & --- \\
    \midrule
    \multicolumn{3}{@{}l}{\textbf{(b) Beam-management thread} ($T_{\text{bm}}$, two UEs, one 50\,ms window)} \\
    \midrule
    Radio-field rendering ($10\times0.996$)      & 9.960 & M \\
    Beam-candidate extraction (GPU)              & 0.205 & M \\
    Backward beam-tracing                        & 0.109 & M \\
    MU-MIMO beam synthesis ($K=2$)               & 0.236 & M \\
    \cmidrule(r){1-3}
    \textbf{Total} $T_{\text{bm}}$               & \textbf{10.510} & M \\
    \midrule
    \multicolumn{3}{@{}l}{\textbf{(c) Motion-to-photon, deployment} ($T_{\text{MTP}}$, 120\,Hz)} \\
    \midrule
    Uplink pose feedback (5G-NR CG)     & 1.500 & B \\
    gNB processing and pose alignment   & 0.500 & B \\
    Server-side XR rendering (GoP 2)    & 16.667 & B \\
    HEVC encoding (NVENC, throughput)   & 2.206 & M \\
    Downlink transmission (5G-NR PHY)   & 1.000 & B \\
    UE demodulation and LDPC decoding   & 0.500 & B \\
    HEVC decoding (hardware, 1 frame)   & 2.520 & M \\
    Display refresh wait (120\,Hz)      & 4.167 & A \\
    \cmidrule(r){1-3}
    \textbf{Total} $T_{\text{MTP}}$     & \textbf{29.060} & --- \\
    \midrule
    \multicolumn{3}{@{}l}{\textbf{(d) Motion-to-photon, prototype test} (software decode)} \\
    \midrule
    Uplink transport (UDP loopback)         & 0.007 & M \\
    Downlink transport (UDP loopback)       & 0.099 & M \\
    Pose $\rightarrow$ decoded (TCP round-trip) & 29.832 & M \\
    Display refresh wait (120\,Hz)          & 4.167 & A \\
    \cmidrule(r){1-3}
    \textbf{Total} $T_{\text{MTP}}$\textbf{, prototype test} & \textbf{33.999} & M \\
    \bottomrule
  \end{tabular}\vspace{-4mm}
\end{table}

\subsection{Latency Analysis}
\label{sec:appendix_latency_analysis}

 To fully contextualize the performance of \sysname, it is essential to distinguish between the beam-management latency ($T_{\text{bm}}$), the scene-update latency ($T_{\text{update}}$), and the motion-to-photon (MTP) latency ($T_{\text{MTP}}$), as they operate on fundamentally different levels and their totals are not additive. Table~\ref{tab:latency_breakdown} reports the full per-stage breakdown of all three concurrent threads. For the two threads that cross the wireless link (scene update and motion-to-photon), we report a \emph{deployment} budget whose network hops take standard 5G-NR air-interface values; panel~(d) reports the motion-to-photon chain on our \emph{current testbed} in which those hops become a host loopback, timed both as a UDP one-way floor and as a live two-process TCP round-trip. Each row is labelled measured (\emph{M}), analytic (\emph{A}), or budgeted (\emph{B}).

Beam management operates in parallel with the lower physical layer and takes place prior to actual frame transmission. Specifically, the system utilizes predicted poses at uniform time steps within a future scheduling horizon ($T_{\text{horizon}}$). Using these poses, \sysname renders beam candidates from the Tx to the Rx, optimizes the global beamforming, and dispatches the corresponding commands to the Rx before the actual transmission occurs. A critical constraint here is that the scheduling horizon must strictly exceed the beam management latency (i.e., $T_{\text{horizon}} > T_{\text{bm}}$). 

For \sysname, the server-side computation for beam management completes in $T_{\text{bm}} \approx 10.5$\,ms for two multiplexed UEs. It is dominated by radio-field rendering of the predicted frames ($10 \times 1.0$\,ms---the five frames of the $50$\,ms window rendered serially for each of the two UEs), followed by beam-candidate extraction ($0.3$\,ms) and MU-MIMO beamforming optimization ($0.2$\,ms), both negligible after moving peak extraction to the GPU. 
This budget is deliberately loose: \sysname's scheme only requires $T_{\text{bm}}$ to finish within one scheduling horizon ($T_{\text{bm}} < T_{\text{horizon}}$), so heavier multi-user loads are absorbed without missing the horizon. The horizon length is the real design knob: a longer $T_{\text{horizon}}$ leaves more room for multi-user scheduling but pushes predictions further into the future and thus degrades dynamic-prediction accuracy, whereas a shorter one improves prediction accuracy yet tightens the scheduling budget. Our $50$\,ms setting is the trade-off we adopt between multi-user scheduling pressure and dynamic-prediction accuracy. Table~\ref{tab:latency_breakdown}(b) lists the per-stage breakdown of this thread. 

On the other hand, MTP latency represents the end-to-end delay experienced by the user. This pipeline spans from the moment the UE uploads its latest update for rendering content, through the server-side rendering of the newest frame, encoding into a Group of Pictures (GoP), transmission, and finally to the UE-side decoding and display. A major determining factor of this latency is the GoP length utilized during rendering. While a longer GoP achieves higher compression rates and reduces bandwidth requirements for a target video quality, it significantly inflates the MTP latency. Because the accurate beam management in \sysname secures high-capacity, high-bandwidth links, the system is no longer bottlenecked by strict bandwidth constraints. This allows \sysname to utilize a short GoP length, which compresses $T_{\text{MTP}}$ 
to $\approx 29$\,ms at 4K 120\,Hz. This budget is primarily dominated by the two-frame render buffer ($2/120 = 16.7$\,ms) and the half-frame display wait. To accurately evaluate the live stream, we measure the codec delays frame-by-frame at 4K resolution. On the server side, hardware encoding (NVENC) takes roughly 2.2\,ms per frame. The primary difference between our ideal deployment budget and our actual prototype lies in the UE decoding stage. While a commercial hardware decoder (NVDEC) processes a frame in just 2.5\,ms (and optimized software decoding takes 3.5\,ms), our commodity prototype relies on single-threaded software decoding, which costs 13\,ms per frame. The remaining stages adopt standard 5G-NR and server-rendering budgets. To validate the entire system, we ran the complete motion-to-photon chain live (Table~\ref{tab:latency_breakdown}(d)), achieving an MTP latency of 34.0\,ms (p99: 40.3\,ms).

Finally, Table~\ref{tab:latency_breakdown}(a) details the scene-update thread, whose $T_{\text{update}} \approx 3.8$\,ms (2.3\,ms of processing plus the 5G uplink hop) keeps the radio twin consistent with the physical scene (\S\ref{ss:dynamic_radio_field}); the prototype test replaces that hop with a 0.007\,ms host socket, giving 2.3\,ms on the bench. This thread is asynchronous: beam management reads the most recently updated scene and never blocks on it, so $T_{\text{update}}$ adds to neither $T_{\text{bm}}$ nor $T_{\text{MTP}}$.

\subsection{Per-Element Array Processing and Cross-Aperture Evaluation}\label{sec:appendix_array_spec}
The public 28\,GHz dataset~\cite{caudill2021real} provides a \emph{per-element} scan of a real $8\times8$ ($64$-element) array, which contains the complex channel response of every element at each Rx position. Because these responses are phase-aligned across elements, any centered $n\times n$ sub-array ($n\le8$) extracted from them is the response of a real physical aperture of that size, obtained without separate captures; a larger aperture combines more elements and yields a narrower main lobe and higher gain. The two-user MU-MIMO scenarios are composed offline from separately measured single-UE traces over this field, with scheduling and null-steering applied in post-processing; phase-coherent dual-UE OTA transmission is left to future work.
\sysname's radio field is trained only on the $4\times4$ sub-array and predicts an aperture-independent beam \emph{direction}, so the same prediction can steer any aperture. We evaluate cross-aperture generalization (Fig.~\ref{fig:mu_mimo_analysis}(a)) by applying that $4\times4$-trained direction to the GT spectrum of the larger sub-arrays and recording the gain lost to its angular offset from the optimum: the same offset costs little on $4\times4$ but more on $8\times8$.

\clearpage

\end{document}

%% file: macros.tex
\renewcommand{\log}{\mathop{\mathrm{log}}}

